\PassOptionsToPackage{dvipsnames}{xcolor}

\documentclass[sigconf]{acmart}

\AtBeginDocument{%
  \providecommand\BibTeX{{%
    Bib\TeX}}}

\usepackage{tikz}
\usepackage{amsmath}
\usepackage{xspace}
\usepackage{tabularx}
\usepackage{multirow}
\usepackage{soul}
\usepackage{enumitem}
\usepackage{listings}
\usepackage{hyperref}
\usepackage{graphicx}
\usepackage{acmart-taps}

\aptLtoXcmd{\newcommand{\textlabel}[2]{}}{\newcommand{\textlabel}[2][Black]{%
  \tikz[baseline=(X.base)]
    \node[rectangle, rounded corners=2pt, fill=#1!25, text=#1, inner xsep=1.5pt, inner ysep=2.5pt] (X) {\texttt{#2}};\xspace%
}}

\definecolor{rblue}{HTML}{0F8CBA}

\newcommand{\system}[0]{{\raisebox{0.25ex}{\scalebox{1}[0.85]{\textsc{J}}}}\textsc{elly}\xspace}
\newcommand{\sys}[0]{{\raisebox{0.25ex}{\scalebox{1}[0.85]{\textsc{J}}}}\textsc{elly}\xspace}
\newcommand{\user}[0]{Millie\xspace}

\aptLtoXcmd{\long\def\sval{{\xbox{aptbox}{\XMLaddatt{style}{background-color:\#dfe9fb;color: \#6495ED;border-radius: 5px; padding:5px}\textsand{SVAL}}}}\xspace}{\newcommand{\sval}{{\textlabel[CornflowerBlue]{SVAL}\xspace}}}

\aptLtoXcmd{\long\def\dict{{\xbox{aptbox}{\XMLaddatt{style}{background-color:\#f6ddf5;color: \#da70d6;border-radius: 5px; padding:5px}\textsand{DICT}}}}\xspace}{\newcommand{\dict}{{\textlabel[Orchid]{DICT}}}}

\aptLtoXcmd{\long\def\pointer{{\xbox{aptbox}{\XMLaddatt{style}{background-color:\#d8ffd8;color: \#00ff00;border-radius: 5px; padding:5px}\textsand{PNTR}}}}\xspace}{\newcommand{\pointer}{{\textlabel[ForestGreen]{PNTR}}}}

\aptLtoXcmd{\long\def\arr{{\xbox{aptbox}{\XMLaddatt{style}{background-color:\#fff1d8;color: \#ffa500;border-radius: 5px; padding:5px}\textsand{ARRY}}}}\xspace}{\newcommand{\arr}{{\textlabel[BurntOrange]{ARRY}}}}

\newcommand{\code}[1]{\texttt{#1}}

\definecolor{hlyellow}{RGB}{255,255,0} 
\sethlcolor{hlyellow}

\colorlet{punct}{red!60!black}
\definecolor{background}{HTML}{EEEEEE}
\definecolor{delim}{RGB}{20,105,176}
\colorlet{numb}{magenta!60!black}

\lstdefinelanguage{json}{
    basicstyle=\normalfont\ttfamily\footnotesize,
    numbers=left,
    numberstyle=\scriptsize,
    stepnumber=1,
    numbersep=8pt,
    showstringspaces=false,
    breaklines=true,
    literate=
     *{0}{{{\color{numb}0}}}{1}
      {1}{{{\color{numb}1}}}{1}
      {2}{{{\color{numb}2}}}{1}
      {3}{{{\color{numb}3}}}{1}
      {4}{{{\color{numb}4}}}{1}
      {5}{{{\color{numb}5}}}{1}
      {6}{{{\color{numb}6}}}{1}
      {7}{{{\color{numb}7}}}{1}
      {8}{{{\color{numb}8}}}{1}
      {9}{{{\color{numb}9}}}{1}
      {:}{{{\color{punct}{:}}}}{1}
      {,}{{{\color{punct}{,}}}}{1}
      {\{}{{{\color{delim}{\{}}}}{1}
      {\}}{{{\color{delim}{\}}}}}{1}
      {[}{{{\color{delim}{[}}}}{1}
      {]}{{{\color{delim}{]}}}}{1}
      {true}{{{\color{numb}{true}}}}{4}
      {false}{{{\color{numb}{true}}}}{5}
}

\AtBeginDocument{%
  \providecommand\BibTeX{{%
    \normalfont B\kern-0.5em{\scshape i\kern-0.25em b}\kern-0.8em\TeX}}}

\copyrightyear{2025}
\acmYear{2025}
\setcopyright{cc}
\setcctype{by-nc-sa}
\acmConference[CHI '25]{CHI Conference on Human Factors in Computing Systems}{April 26-May 1, 2025}{Yokohama, Japan}
\acmBooktitle{CHI Conference on Human Factors in Computing Systems (CHI '25), April 26-May 1, 2025, Yokohama, Japan}\acmDOI{10.1145/3706598.3713285}
\acmISBN{979-8-4007-1394-1/25/04}

\begin{document}

\title[Generative and Malleable User Interfaces with Generative and Evolving Task-Driven Data Model]{Generative and Malleable User Interfaces\\with Generative and Evolving Task-Driven Data Model}

\author{Yining Cao}
\email{rimacyn@ucsd.edu}
\orcid{0000-0002-3962-2830}
\affiliation{%
  \institution{University of California San Diego}
  \city{La Jolla}
  \state{CA}
  \country{USA}
}

\author{Peiling Jiang}
\email{peiling@ucsd.edu}
\orcid{0000-0003-4447-0111}
\affiliation{%
  \institution{University of California San Diego}
  \city{La Jolla}
  \state{CA}
  \country{USA}
}

\author{Haijun Xia}
\email{haijunxia@ucsd.edu}
\orcid{0000-0002-9425-0881}
\affiliation{%
  \institution{University of California San Diego}
  \city{La Jolla}
  \state{CA}
  \country{USA}
}

\renewcommand{\shortauthors}{Cao, et al.}
\begin{abstract}

  Unlike static and rigid user interfaces, generative and malleable user interfaces offer the potential to respond to diverse users’ goals and tasks. However, current approaches primarily rely on generating code, making it difficult for end-users to iteratively tailor the generated interface to their evolving needs. We propose employing task-driven data models---representing the essential information entities, relationships, and data within information tasks---as the foundation for UI generation. We leverage AI to interpret users’ prompts and generate the data models that describe users’ intended tasks, and by mapping the data models with UI specifications, we can create generative user interfaces. End-users can easily modify and extend the interfaces via natural language and direct manipulation, with these interactions translated into changes in the underlying model. The technical evaluation of our approach and user evaluation of the developed system demonstrate the feasibility and effectiveness of the proposed generative and malleable UIs.
\end{abstract}



\begin{teaserfigure}
\centering
  \includegraphics[width=\textwidth]{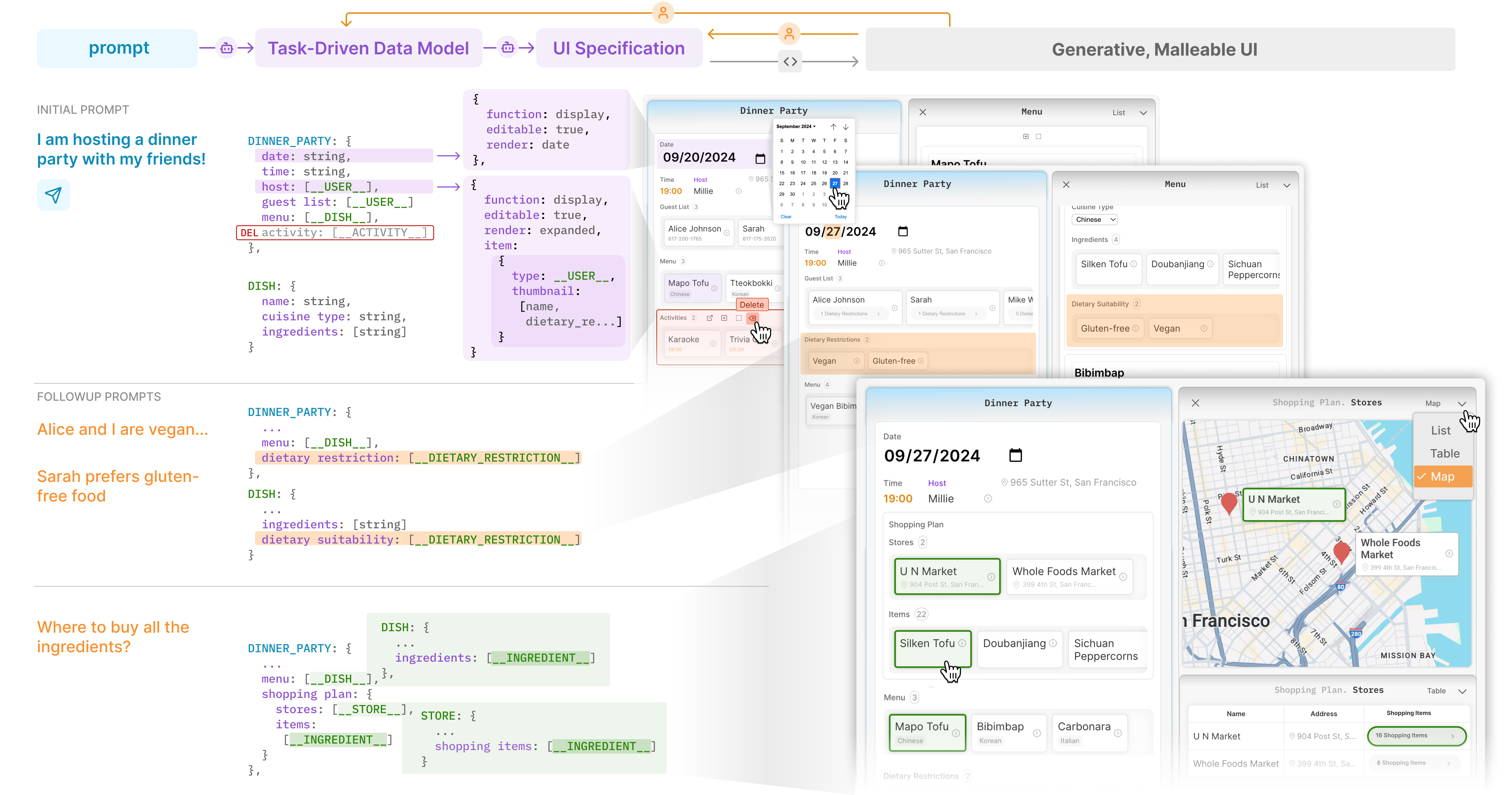}
  \caption{We propose an approach towards generative and malleable user interfaces. With the evolving task-driven data models as foundation, the interface can be generated and dynamically evolves with users' changing information needs.}
  \Description{The figure presents the proposed technical pipeline as a flowchart at the top, illustrating the sequential process. Below the flowchart, three aligned columns visually represent different stages: (1) the user's prompts, (2) the generated task model and corresponding specifications, and (3) the rendered user interfaces. The rendered user interfaces are presenting on three rows, sharing a common structure, with a home panel on the left displaying a "Dinner Party" plan. The right panel varies by row: In the first two rows, it shows a list of menu items. In the third row, it displays a map view.}
  \label{fig:teaser}
\end{teaserfigure}

\maketitle

\section{Introduction}
\label{sec:introduction}

The vision of personalized and intelligent user interfaces, as portrayed in Apple's 1987 Knowledge Navigator \cite{Apple_Knowledge_Navigator}, seems more attainable than ever given the recent advances in AI \cite{GPT4TechReport,anthropic2024claude}. We envision the interfaces to be capable of responding to users' diverse requests, and continuously adapting to users' evolving needs by presenting relevant information with effective representations and interactions. A critical challenge in realizing this vision is devising an interface paradigm and technical approach for creating such generative and malleable user interfaces.

Consider the task of hosting a dinner party. One needs to fix the schedule, invite guests, plan the dishes, compare wine options, finalize a shopping list, and determine the optimal shopping route. Under the dominant application-centric interface paradigm, end-users need to cobble together a large number of applications, using a fraction of the functionality of each to accomplish their goals ~\cite{informationspaceFox, ZhangDataScientistCollaborate}. This fragmented, inefficient workflow is a common experience in our everyday informational tasks. While one could imagine dedicated applications developed for different tasks, it is impractical given the diversity of user needs. In this case, a dedicated ``dinner party'' application may not exist, and if it did, it would likely become bloated with features while still failing to fully accommodate individual preferences and evolving requirements.

The programming capability of generative models offers one approach to achieve generative and malleable UIs: generating the codebase of a custom application from user prompts, which could be compiled and executed to support users' tasks \cite{wu2024uicoder, vercelGenUI, anthropic2024claude}. 
However, the code-generation approach makes it challenging for end-users to modify and extend the interfaces when AI's generation inevitably fails to fully align with users' needs or when those needs naturally shift throughout the task.  
Each new prompt-based revision may result in a discontinuous transition between generated codebases, making it difficult to maintain consistency across iterations; it is also unclear how the data should be transformed when the user's tasks require changing the underlying information structure. 
The opaque relationship between user prompts and the resulting code further complicates interpretability and control, limiting end-users’ ability to steer the generation process effectively.

A potential solution is to introduce higher-level generative structures that can guide both UI generation and data transformations while improving end-user control. This work explores such structures to generate interfaces that are both malleable and interpretable. 

We adopt the canonical perspective that user interfaces are graphical representations of underlying data models that describe the intended tasks, rooted in the Model-View-Controller (MVC) framework for GUI-based applications \cite{mvcdescription} and model-based UI development \cite{puerta1998towards, MyersUISoftware, szekely1996retrospective}. In this view, traditional applications employ \textit{fixed models} for \textit{predefined tasks}, and therefore, the onus is on the users to piece together the separate application models to match their workflows. 
On the contrary, generative and malleable UIs should dynamically evolve to reflect users’ tasks and intentions.
To achieve this, we propose leveraging Large Language Models (LLMs) to interpret users' prompts and generate a \textit{task-driven data model}—a structured representation of the essential entities, relationships, and data properties relevant to the intended task. 
This model serves as the foundation for generating UI specifications that define the components and composition of the interface.
The model will evolve continuously in response to users’ changing needs.
As such, the dynamically generated and evolving task-driven data models can drive the transformation of the interface and the underlying data, achieving generative and malleable UIs. 

This research investigates the feasibility of this approach by exploring both the technical pipeline and interaction techniques of the generative and malleable UIs with a prototype system, ~\sys. The pipeline begins by analyzing user prompts and generating a model that consists of an object-relational schema and a dependency graph. This model then guides the generation of the UIs by representing aspects of the model with predefined UI patterns and rules that reflect common UI design practices \cite{tidwell1997pattern}. Within ~\sys, users can interact with the generated interfaces using natural language and direct manipulation, with these interactions translated to changes in the underlying model. Users can also directly inspect the model to understand the underlying structure of the interface and flexibly customize it to suit their needs. 

We evaluate the LLM-generated data model with a technical evaluation. Results show that state-of-the-art LLMs can reliably generate relevant entities and dependencies that can meet the information and interaction needs expressed in users' natural language prompts. To the best of our knowledge, the generative and malleable UIs we developed are the first of their kind to support relatively open-ended information tasks. To assess their effectiveness, we conducted a user study where participants engaged with the system to complete several open-ended information tasks and reflected on how their experiences compared to those with existing GUI applications and AI-powered chat interfaces. Our findings show that generative and malleable UIs enabled users to develop highly personalized and dynamic information spaces by flexibly curating diverse information and customizing how information is presented.
 
Therefore, this work makes the following contributions: 
\begin{itemize}
    \item A technical approach for generative and malleable UIs based on the generative task-driven data models that evolve with users' tasks.
    \item \sys, a prototype system that implements the proposed approach, enabling users to flexibly generate and customize user interfaces via natural language and direct manipulation.
    \item Technical and user evaluations demonstrating the feasibility of the approach and the strengths and future directions of such generative and malleable interfaces. 
\end{itemize}

\section{Research Framing and Scope}
\label{sec:framingscenario}
Our long-term goal is to develop dynamic, personalized, and adaptive interfaces that cater to individuals' unique and evolving needs across various domains, scopes, and levels of complexity. 
Such interfaces are particularly well-suited for information tasks that require integrating data from multiple domains, presenting information in highly personalized ways, and supporting open-ended or exploratory workflows where users' goals and information needs continuously evolve.
These types of tasks include but are not limited to planning tasks (e.g., travel and party planning), exploring and learning about a new domain (e.g., red wine, cheese), or multi-factor decision-making tasks (e.g., holiday gift shopping, choosing colleges to attend). 

\subsection{Research Scope}
\label{sec: aspects}

For generative and malleable UIs to effectively support users' information tasks with our model-based approach, the following key aspects need to be addressed:
(1) \textit{Task Representation.} The model needs to employ an effective representation to describe users' information tasks;
(2) \textit{UI Generation.} The model needs to be translated into rich and effective interface representations;
(3) \textit{Model Evolution.} The model needs to dynamically adapt to users' evolving tasks and various customization needs.
(4) \textit{Data Integration.} The model needs to be populated with accurate and relevant data.
(5) \textit{Context Awareness.} The interface should incorporate users' context to provide personalized information and UI configurations.

In this paper, we focus on the first three aspects. While fully addressing all aspects is beyond our scope, our technical pipeline and prototype system presented in this work provide a clear guide for future research to continue advancing this paradigm, which we will discuss in detail in Section ~\ref{sec:discussion}. Below, we describe an example scenario of using \system. In this scenario, we assume \system has access to the user's personal information.

\subsection{Envisioned Scenario}

\user is planning to host a dinner party with her friends. Typically, she would have to use the browser to search dishes online, find their recipes, use a note-taking application to record the ingredients and make a shopping plan, use a calendar and several communication applications to coordinate the schedule with her friends, and more. 

Instead of juggling all these applications, she opens \system and types, ``\textit{I am hosting a dinner party}.'' \system responds with a few follow-up questions with generated GUIs, such as ``who to invite'' with a list of selectable contact cards; and ``when the party is'' by presenting a calendar populated with her schedule.
After a brief conversation,
\system generates a home panel for the ``Dinner Party Plan'' task, including the time, location, guest list, menu, and activities. 
Seeing this, \user realizes that she has forgotten to invite a few people. With \sys, she can pull up her contact list by clicking on the ``all'' button beside the guest list. The panel of all her contacts is displayed side-by-side. Contacts already added to the guest list are highlighted in both the guest list and the contact list. Millie then adds the missing guests by tapping on their contact cards.

\user reviews the recommended dishes in the menu, deletes the ones she dislikes, and clicks the ``add'' button next to the menu list to explore more options suggested by \system. While doing so, she realizes that she needs to consider dietary restrictions. She informs \sys, ``\textit{Alice and I are both vegan}.'' To fulfill this request, \sys adds a ``dietary restrictions'' attribute for all guests and automatically records Alice's and Millie's preferences. Meanwhile, a ``dietary suitability'' attribute has been added for each dish, flagging dishes violating the dietary restrictions. \user then replaces them with suggestions made by \sys.
To ensure awareness of the restrictions when planning the activity, \sys also adds a new section to the home panel summarizing the dietary restrictions of all guests. 

When the menu is finalized, \user then types ``\textit{I need to get the ingredients}.'' Recognizing the task shift, ~\system generates a Shopping List panel, organizing the ingredients as shopping items with attributes assisting the ingredient purchasing process: The ``total quantity'' aggregates the amount needed for all the dishes; the ``store'' drop-down menu lists all the local stores where the ingredients are available; and the ``bought'' checkbox tracks the shopping progress. \sys presents the stores in a map view for her to plan the shopping trip. Clicking on each store on the map shows her all the items that she needs to buy at the store. After reviewing the list, she clicks the ``start'' button at the bottom of the map, hops in her car, and heads out to shop.

\section{Related Work}
\label{sec:related}

We adopt the canonical perspective that user interfaces employ interactive graphical representations to encode the underlying data model, similar to the Model-View-Controller (MVC) software design framework for GUI-based applications~\cite{mvcdescription}, 
and the model-based UI development paradigm~\cite{MyersUISoftware, szekely1996retrospective}. 
Taking this perspective, traditional GUI applications employ fixed data models and fixed encodings (i.e., program) to create fixed interfaces, resulting in rigid applications designed for specific tasks with specific features. 
Extensive research has explored various approaches towards the creation of adaptive, dynamic, malleable, and generative UIs. In what follows, we review the key approaches that have been explored.

\subsection{Model-Based User Interfaces} 
Model-Based User Interfaces (MBUI) development arose as a paradigm that aims to significantly reduce the effort in developing UIs while ensuring quality \cite{myersSurveyonUIProgramming, MyersUISoftware}, initiated by early works on User Interface Management System (UIMS)~\cite{olsen1992user, myers1995user} that proposed decoupling application functionality from the UI. MBUI provides a systematic approach to software design by leveraging abstract models, including task models to structure task workflows, domain models to represent data relationships, and abstract-to-concrete mappings to render the UI components. Utilizing these models, developers can specify interfaces declaratively at a higher level of abstraction ~\cite{szekely1996retrospective, klemmer2004tangible}. For example, rather than concretely specifying the UI components, such as a set of radio buttons or a dropdown list, MBUI tools allow developers to define their needs declaratively, such as ``a widget for selecting a single item from a set,'' so the system can decide the most appropriate widget to display based on the specific user scenario, for example, screen sizes of the devices.

Prior works have explored supporting different subprocesses and UI scopes for MBUI. For example, UIDE focuses on dialog box generation by assigning data types to widgets and laying them out on a canvas~\cite{foley1991uide}, while MIKE generates menus and dialog boxes directly from function signatures~\cite{olsen1986mike}. MASTERMIND~\cite{ szekely1996declarative} and ITS ~\cite{wiecha1990its} expand on these approaches by supporting a broader range of interface specifications. Automating the mapping from high-level models to UI specifications has also been a significant focus in prior research. For instance, TRIDENT balances automation with manual refinement, allowing developers to specify presentation and navigation strategies ~\cite{vanderdonckt1995knowledge}. Despite these advancements, one persistent challenge in MBUI is translating the abstract models into concrete UI components. Systems like HUMANOID~\cite{szekely1992facilitating, luo1993management} and BOSS~\cite{schreiber1994specification} use reusable templates to address this issue. TIMM further generalizes these solutions into a computational framework that explicitly represents and manages the mappings between abstract and concrete elements ~\cite{puerta1998towards}.

Our approach takes a similar perspective of separating interface presentation from the underlying system logic that is governed by task-driven data models, and explores automating the process of encoding and mapping between these layers. The primary focus of MBUI, however, has been on assisting developers in creating UIs rather than enabling end-users to modify their interfaces dynamically. Therefore, the models that the MBUI approach produces are predefined and static. Our work, on the other hand, aims to continuously update the underlying model, which drives the transformation of user interfaces to meet the end-users' evolving needs.

\subsection{Specification-Based UI Generation}
Different from the traditional MBUI development paradigm, with which software developers determine the underlying task model, UI model, and their mappings to create a single system \cite{wiecha1990its,puerta1998towards}, the specification-based UI generation can be seen as a scaffolded approach of MBUI such that interfaces can be specified by end-users or automatically generated. This is typically achieved by constraining one or more aspects within the MBUI approach. For example, Bespoke relies on the specification of command-line applications, and by predefining the mappings between UI widgets to different types of command-line parameters, it enables end-users to create GUIs for command-line applications through demonstration \cite{bespoke}. 
Similarly, DynaVis leverages the Vega-Lite specification \cite{vega-lite} to generate visualization editing interfaces. By mapping UI widgets to visualization parameters, it composes appropriate UI components based on parameters inferred from users’ natural language queries \cite{vaithilingam2024dynavis}.
To ensure the quality and consistency of UIs for controlling home appliances, Nichols et al. employed a specification language and parameterized templates that encode design conventions, enabling the automatic generation of structured and coherent interfaces \cite{NicholsSmartTemplate,GeneratingRemoteControl}.

A unique strength of high-level specification is that it enables developers and end-users to focus on composing high-level domain-specific primitives, delegating low-level execution to the underlying architecture and runtime~\cite{heer2010declarative}. By choosing an appropriate level of abstraction and enabling one-to-one mappings between specification and user interface components, end-users can often directly manipulate the specification itself to adjust the generated outcome. 
Additional interface layers, such as graphical or natural language interfaces, can also be utilized on top of the specification if fully instantiating the specification is tedious \cite{vaithilingam2024dynavis, bespoke,narechania2020nl4dv}.

In this approach, domain experts define the specifications, shaping the UI generation space within a structured yet adaptable framework. While high-level specifications impose constraints compared to low-level programmatic approaches, they improve accessibility by making the entire UI generation space more interpretable and modifiable by end-users. Moreover, these constraints help maintain design consistency and quality across generated interfaces \cite{NicholsSmartTemplate}. 
Building upon this approach, we developed a set of UI specifications to translate the data model into interface representations.

\subsection{End-User Programming/Development}
Prior research has explored creating interactive systems that are customizable by end-users, allowing them to tailor the interface to suit their specific needs. This body of work is primarily situated within the field of end-user programming or development (EUP or EUD) \cite{nardiProgramming}, where end-users can employ natural language programming \cite{PUMICE}, GUI-based interaction \cite{litt2020wildcard, Hypercard, fusion}, visual programming \cite{Scratch, DynamicLand}, and programming-by-demonstration \cite{vegemiteJames} to extend existing systems. 

For example, pioneering systems like OpenDoc \cite{Apple_OpenDoc}, HyperCard \cite{Hypercard}, Smalltalk \cite{kay1996early}, and recent systems, such as DynamicLand \cite{DynamicLand} and Embark \cite{sonnentag2023embark} aimed to develop dynamic and personal media for end-users to create their own dynamic content and UIs. Hypercard allows users to develop interactive multimedia content by linking objects via GUI and scripting advanced behaviors using a built-in programming language \cite{Hypercard}. These systems pre-define the underlying data model but expose the encoding mechanism to enable end-user development.

Most existing systems, however, do not expose their data models and encoding mechanisms to the end-users. To circumvent this, research has explored constructing external data models and associated encoding mechanisms to enable EUD. 
For example, Wildcard leverages the accessible and manipulable Document Object Model (DOM) of webpages to enable end-users to collect data from and inject data back into the DOM structure. By representing webpage data in an external spreadsheet, users can manipulate the spreadsheet to customize web pages \cite{litt2020wildcard}.
Further leveraging the composability and transclusion of the web \cite{Nelson1995Transclusion}, prior work explored enabling end-users to create mashup applications tailored to their specific needs. For example, Fusion~\cite{fusion} and C3W~\cite{fujima2004c3w} enable users to create mashups by extracting components from existing webpages and connecting them using transclusion, formula, and glue code. Vegemite allows users to collect data from multiple websites; using scripts that can be generated from users' demonstration, it can perform computation on the collected data and automatically execute web actions such as clicking links and inputting data values to web forms \cite{vegemiteJames}.
In cases where DOM-like accessibility is unavailable,
Research has explored reverse-engineering to extract useful structure and metadata from UI elements. 
For example, Prefab explores recognizing UI widgets on any GUI applications, and then modifying their behaviors using input and output redirection \cite{prefab}. Zhang et al. developed machine learning models to extract metadata from UI screens, which can be used to enhance the screen's accessibility \cite{screenrecogniton}.

While EUP/EUD allows users to extend applications, they lack direct access to the applications' internal data models. 
While external data models (e.g.,  spreadsheet, recognized interface structures) can serve as the proxies or connectors with the original applications, these external data model are also pre-defined by developers, leaving end-users with limited customizability.

\subsection{Context-Based UI Adaptation}
Prior work has also explored computationally adapting UIs based on various contextual factors and constraints pertinent to the device, user, or situation in domains such as accessibility, ubiquitous computing, and mixed reality \cite{dey2001conceptualcontext, abowd1999context, computationalApproachUIGen, wobbrock2011ability, gajos2010personalization}. For example, SUPPLE and SUPPLE++ computationally adjust the size, style, and layout of widgets to adapt the user interfaces based on device constraints (e.g., screen size, input modality) and users' capabilities (e.g., motor and vision) \cite{gajos2004supple, gajos2010personalization, supple++}.   UNIFORM~\cite{uniformNichols} and Huddle~\cite{HuddleAutoGenUIAppliances} automatically generate UIs with primitive widgets for controlling home appliances by considering users' interaction history as well as modeling the similarities and dependencies of appliances. Since mobile and Mixed Reality (MR) interfaces can be invoked in arbitrary situations, research has also explored adapting these interfaces based on various environmental factors \cite{lindlbauer2019context, dey2001conceptualcontext}. For example, Lindlbauer et al. explored how applications should adapt the amount of information they show and their spatial arrangement in MR\cite{lindlbauer2019context}.

With this approach, the data models of the interactive systems are often extended with developer-defined constraints, which will take effects with anticipated contextual input, resulting in context-aware dynamic interfaces. However, the scope of the dynamic behaviors is pre-defined by developers. Therefore, the adaptability---\textit{what} and \textit{how} to adapt---is often less controllable by end-users.

\subsection{AI-based Code Generation}
Recent developments in AI, especially its ability to generate functional program code from natural language prompts, have sparked a new approach towards generative UI. AI products such as Claude \cite{anthropic2024claude} and Vercel \cite{vercelGenUI} can generate and render UI code from natural language prompts. Wu et al. explored fine-tuning LLMs with automated feedback to improve the quality of the generated UI code \cite{wu2024uicoder}. However, they found that the state-of-the-art AI models could struggle to reliably produce compilable programs (less than 80\% compilation success for a single UI screen). 

As AI's code generation capability continues to improve~\cite{anthropic2024claude}, the complexity and quality of AI-generated applications and UIs are expected to increase. This approach presents both opportunities and challenges. On one hand, AI can generate code from arbitrary user requests, making it a more scalable approach for UI generation. On the other hand,  the inherent entry barriers associated with programming languages and tasks as well as the opaque mappings between natural language prompts and generated code create significant challenges for end-users in understanding, controlling and customizing the output \cite{ko2004six, whyjonnycantprompt}.
Given AI’s inconsistent performance---even in generating single-screen UIs---it remains unclear how AI-based code generation can reliably and continuously adapt interfaces to meet users' dynamic and shifting goals. Additionally, current AI-driven code generation approaches primarily focus on client-side UIs, leaving open challenges on how the server-side data should be structured and transformed~\cite{wu2024uicoder, vercelGenUI}.

As a known issue observed across many domains of AI-generated content, creating and iterating via prompting is inherently challenging~\cite{whyjonnycantprompt, sarkar2022like, designguidelinesforpromptengineering, controlnet}. To address this, additional high-level control structures are often required. 
Therefore, it is important to devise high-level structures to guide the generation process, such as imposing additional conditioning constraints for image generation \cite{controlnet} and leveraging compositional structures to ground video generation~\cite{cao2025compositional}. In this work, we take a similar approach by introducing task-driven data models—a high-level control structure that guides UI generation and enables users to more easily inspect and adjust the generated interfaces.

\section{Design Goals}
\label{sec:design}

To achieve the envisioned generative and malleable UIs that can respond to and evolve with users' information tasks, we build upon the existing endeavors and propose the following design goals: 

\begin{itemize}[leftmargin=2.5em]

\item [\textbf{DG1}]  
\textbf{Developing Effective Task-Driven Data Model to Represent Users' Information Tasks}. To support users' information tasks with effective UIs, the underlying foundation--the data model--must be able to effectively represent the information tasks. Unlike some traditional models in MBUI that prescribe the interaction and task sequences \cite{jacobspecification}, which can lead to rigid workflows, we will design the model to represent the essential entities, relationships, and constraints needed to accomplish the information tasks, allowing users to form their own workflows. Additionally, we aim to design the model in a way that it can be intuitively interpreted and manipulated.

\item [\textbf{DG2}]  \textbf{Translating the Task-Driven Data Model into Effective UIs with UI Specification.}
With the model as the foundation, we aim to take the specification-based approach to ensure the consistency and expressiveness of the generated UIs. To effectively map the abstract models to concrete UIs, the specification should be grounded based on a set of common design patterns that describe what UI widgets should be used and how they should facilitate interaction with different types of information. We aim to use generative AI for generating the specifications; therefore, the specifications should be designed to be effectively leveraged by AI for robust and accurate generation of user interfaces.

\item [\textbf{DG3}]  \textbf{Providing Interactions for End-Users to Modify the UIs to Align with Their Evolving Needs.} 
To accommodate various interaction modalities and levels of specificity, we aim to empower end-users to express their intended tasks and UI modifications through both natural language prompts and direct manipulation. These interactions will be translated into updates to the underlying model. In addition, we plan to provide an “Inspect”-like tool, similar to browser developer tools, for end-users to directly examine and edit the model for enhanced interpretability and control over both the generation process and resulting interfaces.
\end{itemize}

\section{Technical Pipeline for Generative and Malleable User Interface}
\label{sec:pipeline}

\begin{figure*}[!ht]
\centering
\includegraphics[width=1\textwidth]{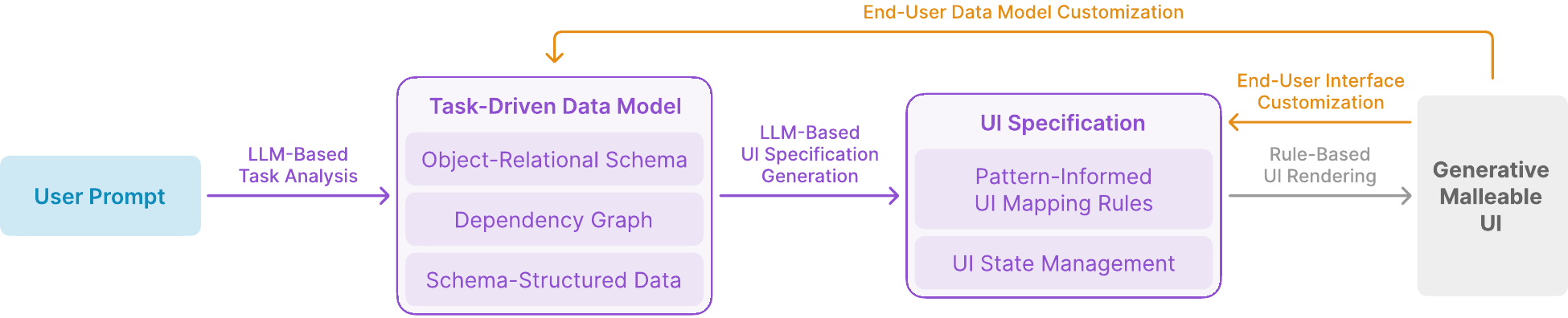}
\caption{Our proposed pipeline for generative and malleable UIs. The pipeline takes the user prompt as input and employs LLMs to generate a data model describing the task. This model serves as the foundation for generating UI specifications, which guide UI composition and state management. The specifications are then used to render the UI based on predefined rules. Users can iteratively customize both the data model and UI specifications through interactions.}
\label{fig:pipeline}
\Description{A flow chart describing the technical pipeline for Generative and Malleable UIs. The pipeline takes users' prompts as input and employs LLMs to generate the task-driven data model, including object-relational schema, dependency graph, and schema-structured data. The pipeline then generates UI specifications, including pattern-informed UI mapping rules and UI state management. The Jelly UI is then rendered and enables the user to continuously customize the data model and UI specification.}
\end{figure*}

Building upon the design goals outlined above, we propose a technical pipeline that takes users' prompts as input and generates corresponding user interfaces, 
as illustrated in Fig.~\ref{fig:pipeline}. The pipeline begins by analyzing user prompts to infer user goals and derive sub-tasks. This information is then leveraged by LLMs to generate the \textbf{Task-Driven Data Model}, which represents the structure of the task. The data model is then translated into a \textbf{UI Specification} that defines the composition of UI elements and manages their states. Users can continuously provide natural language prompts and directly manipulate the generated interfaces, which are both translated into corresponding changes to the underlying data model, and subsequently drive real-time updates to the underlying data and/or UIs, resulting in generative and malleable user interfaces. We detail each component of the pipeline in the following sections.

\subsection{Task-Driven Data Model Generation}
\label{pipeline:schema}

The task-driven data model comprises three components: (1) the \textit{Object-Relational Schema} that describes the types of entities required by the task, as well as their attributes and relationships; (2) the \textit{Dependency Graph} that describes additional dependency relationships across entities, and (3) \textit{Structured Data} that instantiates the schema and dependencies with concrete values.

\subsubsection{Object-Relational Schema}
We model the user task with an object-relational schema, with which the task and its entities are represented as objects with attributes, and relationships among entities are modeled as references among the objects. Fig.~\ref{fig:schema} shows a sample schema generated with the prompt---``\textit{give me a weekly meal plan}.'' A schema consists of the following elements:

\vspace{2mm}

\noindent\textbf{Task.} The task object is the root of the object-relational schema, which describes the attributes essential to the overall task. For example, the task object of a travel planning task might include attributes such as destination, duration, 
and itinerary. A meal plan task object might include start/end dates and a daily plan (Fig.~\ref{fig:schema}a).

\vspace{2mm}

\noindent\textbf{Entity.} The schema contains entities that model the essential components of a task. For example, the task of creating a meal plan consists of entities such as daily meal plans, recipes, ingredients, and grocery stores (Fig.~\ref{fig:schema}b). In another case,  a literature review task might include entities such as paper and author. Each entity contains attributes and cross-references with other entities. 

\vspace{2mm}

\noindent\textbf{Attribute.} Attributes of the task and entity objects are rendered based on their data types, which can be one of the four types:

\noindent \sval is a singular data value, such as date, location, etc. (Fig.~\ref{fig:schema}c$_{1}$).

\noindent \dict is a dictionary that stores key-value pairs, such as the nutrition facts for a dish entity (Fig.~\ref{fig:schema}c$_{3}$).

\noindent \pointer is a reference to another entity, such as a pointer to a ``store'' entity in a shopping item (Fig.~\ref{fig:schema}c$_{4}$). 

\noindent \arr is a collection of items of \sval or \pointer type (Fig.~\ref{fig:schema}c$_{2}$). Note that schema syntax does not allow array of \dict. If there are multiple entities that share the same \dict, they will be abstracted as an entity, and their references will be treated as \pointer. This abstraction simplifies the data model and ensures consistency in how entities and attributes are handled across the system. 

\subsubsection{Dependency Graph}
\label{sec:dep}
Dependencies are an essential aspect of complex tasks, which manifest in the UI as relationships between components.  Our pipeline uses LLMs to generate these dependencies based on the characteristics of the task,  expressed as: 

\vspace{-10pt}

\begin{equation}
\textit{Dependency} := \{\textit{Source}, \textit{Target}, \textit{Mechanism}, \textit{Relationship}\}
\end{equation}

\textit{Source} and \textit{Target} refer to specific entities or attributes within the object-relational schema.

\textit{Mechanism} defines how the target reacts to the changes of the source in one of the following two ways:
\texttt{Validate} ensures constraints are upheld. For example, the checkout date must be later than the check-in date. If violated, the update of the checkout date value will be rejected, and the UI will highlight the issue to explain the violation to the user.
\texttt{Update} automatically propagates changes. For example, the total calories of a dish update automatically if the quantities of the ingredients change.

\textit{Relationship} defines the relationship between \textit{Source} and \textit{Target}. A JavaScript snippet will be generated if the dependency can be expressed by code, e.g., numerical calculations or validations. Otherwise,  the relationship is described in natural language, which LLMs can process to apply the effects. 

\subsubsection{Structured Data.}

Once the schema and dependency graph are defined, the next step is to acquire data that conforms to the specified structure and constraints. 
The pipeline is designed to support real-time data integration from multiple sources, handling both structured and unstructured data, such as generated data, user-uploaded data, and external APIs (e.g., TripAdvisor for travel~\cite{TripAdvisorAPI} or Semantic Scholar for research~\cite{SemanticScholarAPI}). Since our focus is on evaluating the generation of the data model and the user interface, the current pipeline primarily relies on generated data. However, future work could extend its capabilities to incorporate live data acquisition, which we further discuss in Section~\ref{sec:data}.

\begin{figure*}[t]
\centering
\includegraphics[width=\textwidth]{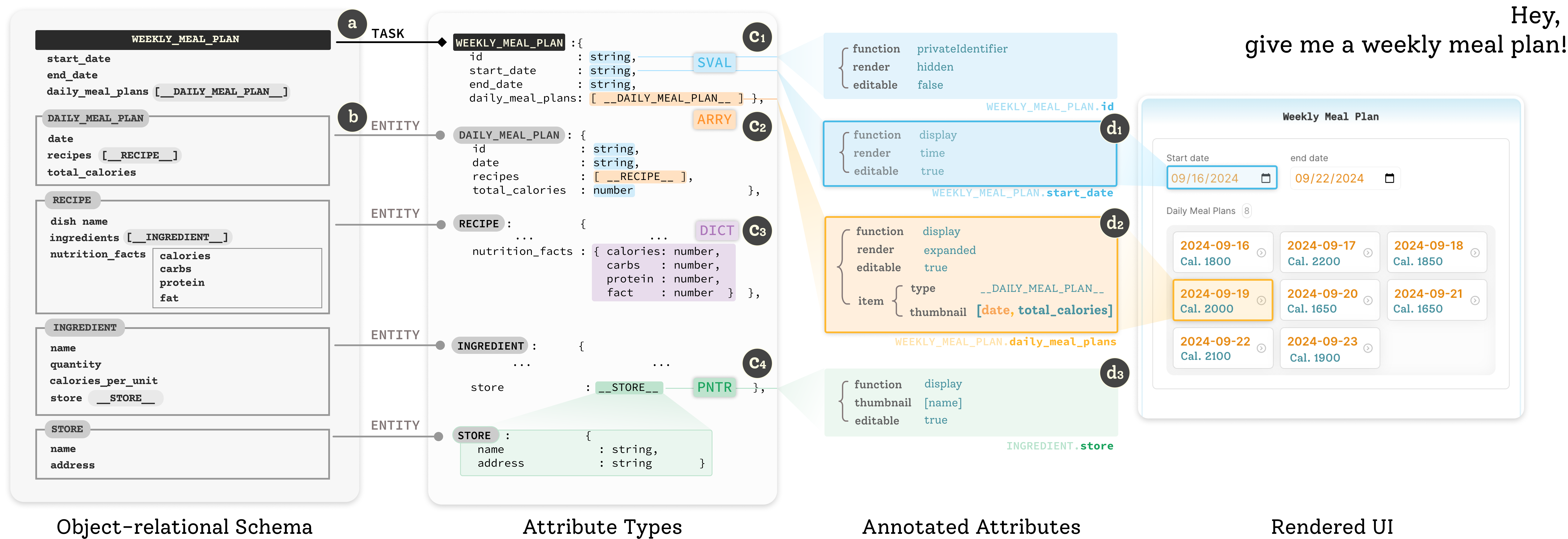}
\caption{The format of object-relational schema and the annotation of attributes within the schema. (a) task attributes of the schema, (b) entities described in the schema; (c1) single data value attribute, (c2) array attribute, (c3) dictionary attribute, (c) pointer attribute; (d1) annotated single data value, (d2) annotated array attribute, (d3) annotated pointer attribute }
\label{fig:schema}
\Description{This figure is composed of four parts placed from left to right. The first part (Object‐Relational Schema) shows a box labeled WEEKLY_MEAL_PLAN with fields such as start_date and daily_meal_plans. These reference other entities named DAILY_MEAL_PLAN, RECIPE, INGREDIENT, and STORE, each with their own attributes (like "dish name" or "nutrition_facts"). The second part (Attribute Types) presents color‐coded examples of single data values (SVAL), arrays (ARRAY), dictionaries (DICT), and pointers (PNTR). The third part (Annotated Attributes) illustrates how each attribute is annotated into UI specifications by defining its function, render types and whether editable. The fourth part (Rendered UI) shows a "Weekly Meal Plan" dashboard UI where users can select a start and end date and see daily meal items as a grid of cards showing date and total calories.}
\end{figure*}


\subsection{UI Specification Generation}
With the task-driven data model, the pipeline then generates the user interface based on the model. To ensure consistency, stability, and quality of the generated UIs, we opted for a specification-based approach explored in prior work \cite{vaithilingam2024dynavis, bespoke, vega-lite}. Therefore, this step of the pipeline takes the data model and generates the UI specification, which will guide the composition of the user interface. 

\subsubsection{Annotating Object-Relational Schema with UI Mapping Rules}
\label{sec:annotations}

Specifically, the pipeline examines each task and entity attribute and annotate each with labels that specify their data types, function roles, and rendering types.
The annotations serve as a specification that guides the mapping of schema elements to UI components using a rule-based approach. We provide the full specification and an example in Appendix~\ref{sec:specs}.

Attributes of \dict, \pointer, \arr may take significant screen space if fully rendered. Therefore, care needs to be taken to ensure the appropriate amount of information is presented on the interface through appropriate UI composition to enable progressive disclosure. Below, we describe how each type of attribute will be labeled and how the labels will affect view composition.

\, \sval is labeled with \texttt{<function, render, editable>} to describe the functional role, the corresponding rendering widget type (e.g., text, time, or location), and if the user may change the value within the GUI.
For example, in Fig.~\ref{fig:schema}d$_{1}$, the \texttt{start\_date} is labeled as \texttt{<display, time, true>}, which will be rendered as a calendar widget on the interface that can receive user edits.

\, \dict itself is not labeled, but all attributes within it will be labeled and directly rendered within \dict attribute's parent view.

\, \pointer is labeled with \texttt{<function, thumbnail, editable>} with \texttt{<function, editable>} the same as \sval. \texttt{<thumbnail>} specifies the attributes in the referred entity that should be displayed for each minimized item. For example, in Fig.~\ref{fig:schema}d$_{3}$, the \texttt{store} attribute for every ingredient is a pointer to a store entity. When rendering the ingredient item on the UI, it will only show the name of the store as a hyperlink to the full details of the store it points to.

\, \arr is labeled with \texttt{<function, render, editable>}. The \texttt{render} type for an \arr can be ``expanded'' or ``summary''. When labeled with ``expanded'', the list will be fully rendered.
When labeled with ``summary'', the list will be rendered in a minimized format, only showing the designated summarizing text and corresponding value, e.g., ``Total Calories \code{2100}'' for a list of dishes. The summarized form, upon clicked, can expand and show the full list. We further illustrate the interaction mechanisms used for navigating these collections of objects in Section~\ref{sec:view_arr}.

\subsubsection{Executing Dependency Graph with UI State Management}

The generated dependency mechanisms are executed with corresponding UI state management rules, ensuring that \sys reliably and consistently handles the logic for the generated UI. The state management unit of the system sandboxes each dependency execution to limit its effects to ensure UI stability, interprets, and executes the updating or validating mechanisms accordingly.

\subsubsection{UI Rendering}

With the specification, the UI rendering process starts with the object-relational schema for the overall task and recursively renders each referred entity and its attributes. Mapping from the specification labels to the UI widgets is handled through a predefined set of rules that ensures consistency across model and data updates. We included the mapping rules used in the current prototype in the supplemental materials.

\subsection{Customization with Continuous Prompting}
\label{sec:cont-prompt}

As users provide follow-up requests to \sys, the pipeline dynamically updates both the data model and the UI. The system leverages previous prompts and data models as context, querying the LLM to determine the necessary update operations. It first assesses whether the request requires modifications to the schema (e.g., adding, removing, or updating entities or attributes) and/or updates to the data. These requests are then parsed into a sequence of operations specified as:

\vspace{-10pt}

\begin{equation}
    \textit{Updater} := \{\textit{Target}, \textit{Action}, \textit{Specifications}\}
\end{equation}

The \textit{Target} refers to the path of the relevant entities or attributes. \textit{Action} includes operations such as add, remove, and update (for both schema and data), as well as data-specific operations like cluster, filter, and sort. \textit{Specifications} details the specific changes to be made for the given action, such as the name of the attribute to be added to the target entity schema.
Based on this, the LLM generates the necessary operations to update the UIs.


\begin{figure*}[ht]
    \centering
    \includegraphics[width=\textwidth]{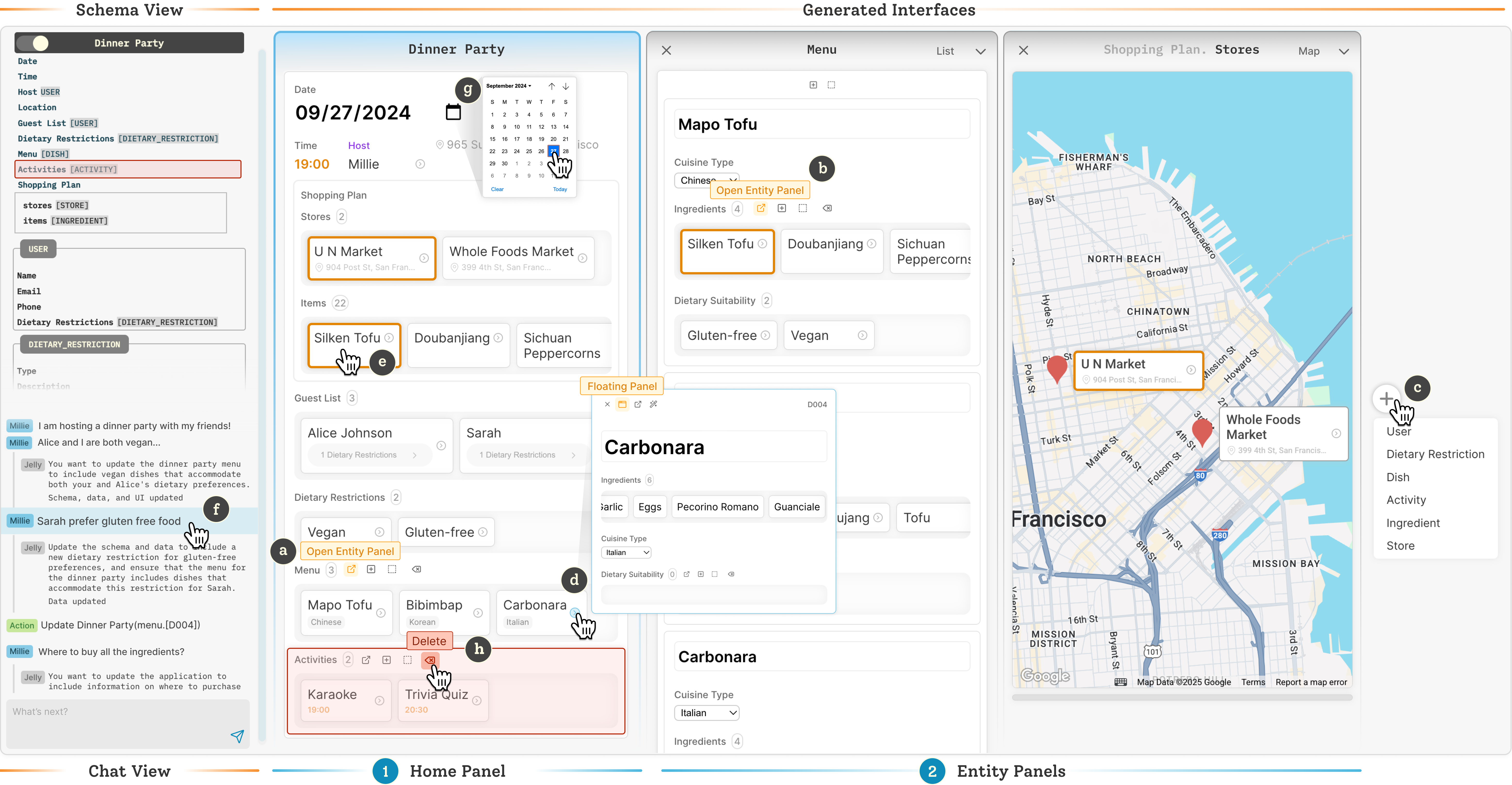}
    \caption[]{\sys consists of the Schema View, the Chat View, and the Generated Interfaces. The task object is presented as the Home Panel (1). Users can navigate to Entity Panels (2) by expanding items in other panels (a-b) or using the navigation button on the right side of the interface (c). Users may click to reveal details of an item (d) and hover over an item to trigger synchronized highlighting across panels (e). Users can customize the UIs and data through continuous prompting (f) and direct manipulation (g-h), with all actions preserved as a traceable history in the Chat View.}
    \label{fig:jellyui}
 \Description{The figure shows the system’s structure, with three main components: Schema View on the left (displaying a ``Dinner Party'' schema and form fields), a Chat View below it (showing user and system messages), and Generated Interfaces in the center/right.  In the central Home Panel, the dinner party’s date, time, guest list, menu items, and stores (``U N Market,'' ``Whole Foods Market'') are displayed. On the right, Entity Panels include a map with store pins and a ``Carbonara'' floating panel. Various icons and labels (a - h) indicate interactions: expanding specific items using specific widgets located next to the rendered item name (a - b) or clicking a dedicated navigation button located at the right of the last panel (c). Interactive Jelly Components enable users to click for details (d) and hover to activate synchronized highlighting across related panels (e). Customization is supported through continuous prompting (f) and direct manipulation (g, h) , ensuring adaptability. A Chat View maintains a history of user interactions and modifications with different tags, including Millie (the user), Jelly (the system), and Action (the translated description of user interaction for modifying the UIs).}
\end{figure*}

\subsection{Implementation} 

Our current pipeline is developed using Python, with the front-end developed in JavaScript using the React framework. To optimize performance, we process independent pipeline requests to LLMs concurrently. For the generation steps, we leverage Anthropic's Claude 3.5 Sonnet and OpenAI's GPT-4o models, selected based on internal performance testing across various pipeline tasks. At each generation step, we incorporate the user's previous request as context. The LLMs are instructed to first infer tasks implied by the user's current prompt, then generate a JSON object in a specified response format. To ensure controlled and accurate outputs, we use few-shot prompting tailored to each step. Additionally, the pipeline performs compatibility checks on the generated schema and data before rendering.
Full implementation details and service usage are provided in the supplemental materials.

\section{\textsc{Jelly}: a Generative and Malleable Interactive System}
\label{sec:interface}

\begin{figure*}[!t]
    \centering
    \includegraphics[width=\textwidth]{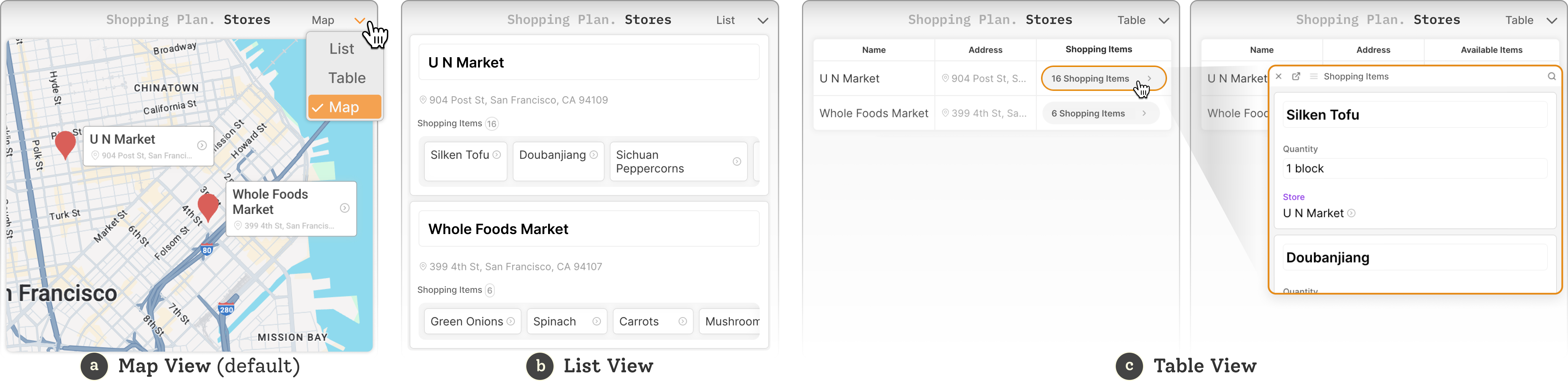}
    \caption[]{\sys currently supports switching between three different representations of data for an entity pane, including (a) \emph{Map View}, (b) \emph{List View} (where each item is expanded to show more details, e.g., the Shopping Items attribute is directly rendered for people to directly scroll through the list), and (c) Table View (where attributes may be folded to fit in the cells, e.g., the Shopping Items list is folded into a button. The user may click on it to reveal the full list).}
    \label{fig:switchrep}
    \Description{The figure shows four side-by-side screenshots demonstrate how the same store information can be viewed in different representations in Jelly, from left to right: (a) map view, a map of San Francisco with pins marking ``U N Market'' and ``Whole Foods Market''.  (b) list view, stores appear as a list of cards showing its name, address, and a scrollable list of shopping items underneath; (c) table view, consists 2 screenshots, on the left stores appear in a table with columns for Name, Address, and Shopping Items; the items themselves are initially folded under a button with summary ``16 Shopping Items'', which can be clicked to open a pop-up panel revealing the full item list, shown on the right. }
\end{figure*}

\sys is a prototype system developed using the technical pipeline described above (Fig.~\ref{fig:jellyui}). Users enter \sys with a prompt that specifies their task. In addition to the main generated interfaces, \sys's sidebar comprises a Schema View, which surfaces the object-relational data schema; and a Chat View that allows users to provide follow-up prompts, where \sys responds in natural language about interface changes. In the following sections, we describe the interface designs and interaction techniques in \sys for effectively supporting users' tasks and customization of the generated UIs.

\subsection{View Management}

User tasks are often supported by complex data models, therefore, presenting all information in a single view can be overwhelming. \sys employs a set of view management strategies that help users comprehend, navigate, and interact with the generated interfaces.

\subsubsection{Panels for Showing Task and Entity Objects}
\label{sec:view_homepanel}
The initially generated interface displays only the Home Panel (Fig.~\ref{fig:jellyui}-1), which corresponds to the top-level task object. This provides users with a clear entry point, offering an overview of the task structure. 

Each entity within the data model is represented by its own panel (Fig.~\ref{fig:jellyui}-2). Users can navigate to these panels to focus on specific aspects of a task by clicking on the \raisebox{-0.2\height}{\includegraphics[height=1em]{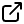}} icon next to the entity name (Fig.~\ref{fig:jellyui}a). 
Entity Panels can also be opened within another Entity Panel when there are references between entities (Fig.~\ref{fig:jellyui}b). 

Through our use of \sys, we recognized that some entities can be deeply nested within others. For example, \texttt{Dietary Restrictions} is not originally displayed in the home panel. Therefore, retrieving all \texttt{Dietary Restrictions} requires navigating through the \texttt{Dish} entity panel first, making accessing and editing these entities cumbersome. To address this, \sys allows users to view all entities and open their respective panels directly using the \raisebox{-0.2\height}{\includegraphics[height=1em]{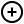}} button (Fig.~\ref{fig:jellyui}c), reducing unnecessary navigation steps. 

Additionally, panels can be closed, resized, or rearranged, allowing users to customize their workspaces as needed.

\subsubsection{Organizing Collections of Objects within Panels}
\label{sec:view_arr}

Given the nested structure of the object-relational schema in our underlying data model, it is important to effectively show collections of objects in the panels. Our design goal is to support users in efficiently navigating complex data structures while maintaining an overview of key information. As described in Section~\ref{sec:annotations}, we support two rendering types to achieve this: 

\textit{Expanded} rendering presents a full list of items, with each item displaying a subset of attributes most relevant to the task. For instance, in the \texttt{Dinner Plan} panel, the \texttt{Menu} attribute is an array of \texttt{Dish}, displaying as a list of items on the interface, with each item only showing its name and cuisine. Clicking on an item opens a popup card, showing full detailed information, such as ingredients and dietary suitability of the dish (Fig.~\ref{fig:jellyui}d). We choose the popup as a default form of revealing the details for in-situ inspection. Alternatively, users can choose to convert the popup card into a persistent floating card with \raisebox{-0.2\height}{\includegraphics[height=1em]{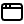}} for easy reference; or open it in a dedicated entity panel for a more focused view.

\textit{Summary} rendering condenses the collection into a single summary button, showing only the most relevant information for the task at hand.  For example, in a \texttt{Shopping Plan}, a list of shopping items can be represented as a button showing the total number of \texttt{Shopping Items}, which users can click to reveal and expand into a full list (Fig.~\ref{fig:switchrep}c). Similarly, within a \texttt{Travel Plan}, the \texttt{Budget} may be summarized as the total sum of all expenses, with an option to click and reveal a detailed breakdown.

\begin{figure*}[!t]
    \centering
    \includegraphics[width=0.8\textwidth]{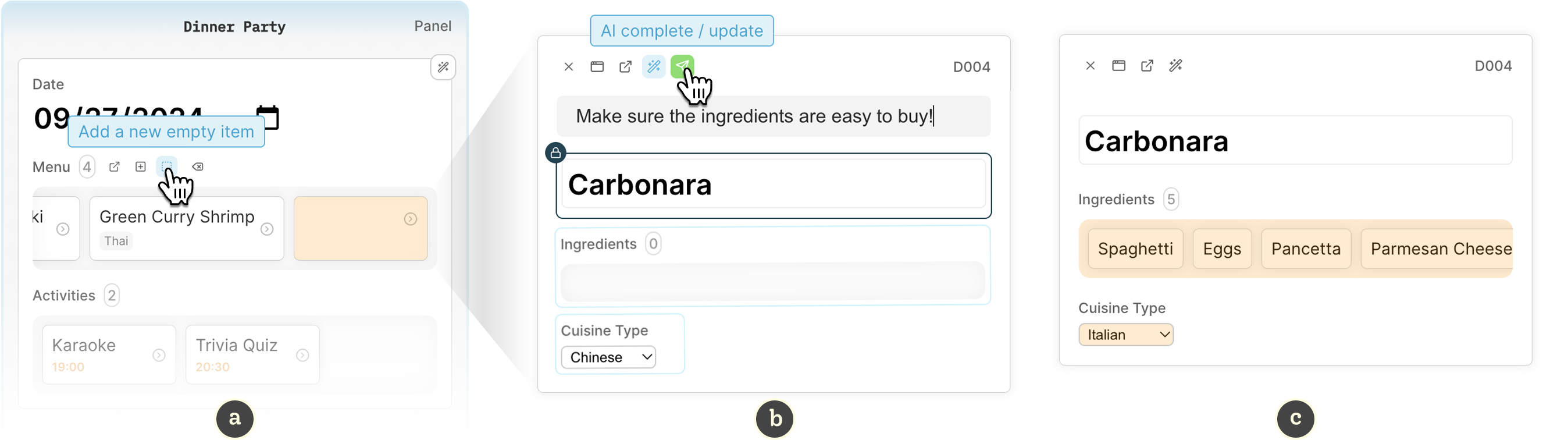}
    \caption[]{Generating Structured Data with \sys. Users can create an empty object card (a), manually input some of the attributes, and optionally provide a prompt specifying their preference over other attributes (b). \sys then auto-completes the remaining attributes based on the provided input (c).}
    \label{fig:autocomp}
    \Description{Three figures illustrating the auto-complete feature in Jelly. For left to right are three screenshots of part of the interface: (a) The user clicks on the square button in the widget bar next to the rendered attribute name to add a new empty item; (b) an empty card pops up, where the user puts the name of the dish, ``Carbonara'', and the prompt, ``Make sure the ingredients are easy to buy!''; (c) The LLM auto-completes the card, with updated values, the ingredients of the dish and the cuisine type, highlighted.}
\end{figure*}

\subsubsection{Cross-referencing with Synchronized Highlighting}
As mentioned above, an entity object can have multiple distinct representations within the interface. For example, an \texttt{Ingredient} object may appear in the home panel, and as an item in both the \texttt{Store} and \texttt{Menu} panel (Fig.~\ref{fig:jellyui}e). 
To facilitate cross-referencing across different views, \sys implements synchronized highlighting. When users hover over one object, all other objects containing the same instance are highlighted simultaneously. This helps users quickly identify related information across different contexts.

\subsection{Interaction Techniques for Customization }

\sys allows users to customize the generated UIs with both natural language and direct manipulation, accommodating different types of customization needs. 

\subsubsection{Continuously Prompting with Traceable History}
Users can give follow-up prompts in the chat view to continuously update the data model and the rendered UIs (Section~\ref{sec:cont-prompt}). Each prompt also serves as an interactive history entry, as it preserves the state of the data model and UI specifications. Users can easily revisit any previous workspaces by clicking on corresponding messages (Fig.~\ref{fig:jellyui}f). Additionally, any user customization made through the GUI, as described in the following section, is translated into an action-tagged entry. With the traceable history, users can easily switch between different versions of the interface geared towards specific tasks, or revert any changes if there are adjustments that do not meet their expectations.

\subsubsection{GUI-Based Data Model Customization}

While continuous prompting enables users to issue high-level requests and make complex structural changes, \sys also provides GUI-based direct manipulation for more granular customization of both data and schema elements.

\textit{Data Customization.}
Data in \sys are editable with suitable representations (Fig.~\ref{fig:jellyui}g). Additionally, the object-oriented underlying structure enables encapsulated actions on entity instances. 
Users can generate additional instances of an entity (e.g., adding more dishes to a menu) by clicking the \raisebox{-0.2\height}{\includegraphics[height=1em]{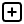}} button. Alternatively, they can add empty instances by clicking \raisebox{-0.2\height}{\includegraphics[height=1em]{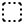}} (Fig.~\ref{fig:autocomp}a). This allows them to fill in the values manually.
In many cases, users may only know partial attributes of an entity. For example, Millie wants to add \texttt{Carbonara} as a dish for the dinner party, but she does not know the ingredients of it. To support this common need, \sys provides an auto-complete feature: clicking the \raisebox{-0.2\height}{\includegraphics[height=1em]{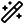}} button allows the user to automatically fill in missing attributes, which also triggers a prompt box at the top of the card, allowing users to specify preferences for the generated attributes (Fig.~\ref{fig:autocomp}b--c).

\textit{Schema Customization.}
Beyond data customization, \sys allows users to directly delete unnecessary attributes using \raisebox{-0.2\height}{\includegraphics[height=1em]{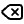}} button next to the attribute name (Fig.~\ref{fig:jellyui}h). However, adding or modifying attributes currently requires using the continuous prompting in the chat view. The implemented interactions represent only a subset of the possible customization techniques that could be integrated into \sys. Given the proposed data model and UI specifications, we anticipate multiple interaction techniques for malleable UIs can be applied here. For example, users could customize which attributes to display in an expanded list when rendering a collection of objects (Section~\ref{sec:view_arr}), as explored in recent research on malleable overview-detail interfaces~\cite{odi}.

\subsubsection{Switching between Representations of Data}
Even when using the same schema and data, users may prefer different representations depending on specific tasks. For example, a \textit{list} facilitates browsing, such as viewing a set of places to visit; a \textit{map} visualizes spatial relationships for route planning; a \textit{table} makes it easier to compare attributes across items for decision makings.

To accommodate this, \sys allows users to flexibly switch between representations within an entity panel, which displays multiple instances of an entity. \sys automatically selects the most suitable representation based on the task, and provides a dropdown menu at the top-right of the panel for users to switch at any time (Fig.~\ref{fig:switchrep}). While the current implementation only supports list, table, and map views, \sys's infrastructure allows for easy extension to include other representations, such as timelines, stacks, or even user-defined representations tailored to specific needs.

\section{Technical Evaluation}
\label{sec:tech}

\aptLtoX[graphic=no,type=html]{\begin{table*}
\caption{Entity and Attribute Coding Results, rated on a 4-point scale, 1 - not relevant or redundant, 2 - could be useful, 3  - necessary and expected, 4 - useful and surprising }
\label{table:dep-1}
\begin{tabular}{cccccccccc}
\toprule
\multicolumn{2}{c|}{} &
\multicolumn{4}{c|}{\textbf{Less Detailed Prompts}} & 
\multicolumn{4}{c}{\textbf{More Detailed Prompts}}\\  
\midrule
{\textbf{Entity}} & \textbf{Total} & \multicolumn{4}{c|}{102} & \multicolumn{4}{c}{95}\\  
\cline{2-10}
& \textbf{Rating}  & \textbf{1} & \textbf{2} & \textbf{3} & \textbf{4} & \textbf{1} & \textbf{2} & \textbf{3} & \textbf{4}\\  
\cline{2-10}
& C1  & 0\% & 3.92\% & 96.08\% & 0\% & 0\% & 4.21\% & 95.79\% & 0\%\\  
& C2  & 0\% & 2.94\% & 92.16\% & 4.90\% & 3.16\% & 2.11\% & 93.68\% & 1.05\%\\  
\cline{3-10} 
& \textbf{Mean} & \textbf{0\%} & \textbf{3.43\%} & \textbf{94.12\%} & \textbf{2.45\%} & \textbf{1.58\%} & \textbf{3.16\%} & \textbf{94.74\%} & \textbf{0.53\%}\\  
\hline
{\centering\textbf{Attribute}} & \textbf{Total} & \multicolumn{4}{c|}{534} & \multicolumn{4}{c}{518}\\  
\cline{2-10}
& \textbf{Rating}  & \textbf{1} & \textbf{2} & \textbf{3} & \textbf{4} & \textbf{1} & \textbf{2} & \textbf{3} & \textbf{4}\\  
\cline{2-10}
& C1 & 0.37\% & 5.81\% & 93.63\% & 0.19\% & 0\% & 0.97\% & 94.59\% & 0.97\%\\  
& C2 & 0\% & 2.62\% & 94.19\% & 3.18\% & 0.97\% & 1.93\% & 95.75\% & 0.77\%\\  
\cline{3-10} 
& \textbf{Mean} & \textbf{0.19\%} & \textbf{4.22\%} & \textbf{93.91\%} & \textbf{1.69\%} & \textbf{0.49\%} & \textbf{1.45\%} & \textbf{95.17\%} & \textbf{0.87\%}\\  
\bottomrule
\end{tabular}
\end{table*}}{\begin{table*}[h]
\centering
\captionsetup{justification=centering}
\caption{Entity and Attribute Coding Results, rated on a 4-point scale, 1 - not relevant or redundant, 2 - could be useful, 3  - necessary and expected, 4 - useful and surprising }
\label{table:dep-1}

\small
\centering
\begin{tabularx}{\textwidth}{>{\centering\arraybackslash}X|
>{\centering\arraybackslash}X|
>{\centering\arraybackslash}X>{\centering\arraybackslash}X>{\centering\arraybackslash}X>{\centering\arraybackslash}X|
>{\centering\arraybackslash}X>{\centering\arraybackslash}X>{\centering\arraybackslash}X>{\centering\arraybackslash}X}
    \toprule
    \multicolumn{2}{c|}{} &
    \multicolumn{4}{c|}{\textbf{Less Detailed Prompts}} & 
    \multicolumn{4}{c}{\textbf{More Detailed Prompts}}\\  
     \toprule
     \hline

\multirow{5}{=}{\centering\textbf{Entity}} & \textbf{Total} & \multicolumn{4}{c|}{102} & \multicolumn{4}{c}{95}\\  
\cline{2-10}
& \textbf{Rating}  & \textbf{1} & \textbf{2} & \textbf{3} & \textbf{4} & \textbf{1} & \textbf{2} & \textbf{3} & \textbf{4}\\  
\cline{2-10}
& C1  & 0\% & 3.92\% & 96.08\% & 0\% & 0\% & 4.21\% & 95.79\% & 0\%\\  
& C2  & 0\% & 2.94\% & 92.16\% & 4.90\% & 3.16\% & 2.11\% & 93.68\% & 1.05\%\\  
\cline{3-10} 
& \textbf{Mean} & \textbf{0\%} & \textbf{3.43\%} & \textbf{94.12\%} & \textbf{2.45\%} & \textbf{1.58\%} & \textbf{3.16\%} & \textbf{94.74\%} & \textbf{0.53\%}\\  

\hline
\hline

\multirow{5}{=}{\centering\textbf{Attribute}} & \textbf{Total} & \multicolumn{4}{c|}{534} & \multicolumn{4}{c}{518}\\  
\cline{2-10}
& \textbf{Rating}  & \textbf{1} & \textbf{2} & \textbf{3} & \textbf{4} & \textbf{1} & \textbf{2} & \textbf{3} & \textbf{4}\\  
\cline{2-10}
& C1 & 0.37\% & 5.81\% & 93.63\% & 0.19\% & 0\% & 0.97\% & 94.59\% & 0.97\%\\  
& C2 & 0\% & 2.62\% & 94.19\% & 3.18\% & 0.97\% & 1.93\% & 95.75\% & 0.77\%\\  
\cline{3-10} 
& \textbf{Mean} & \textbf{0.19\%} & \textbf{4.22\%} & \textbf{93.91\%} & \textbf{1.69\%} & \textbf{0.49\%} & \textbf{1.45\%} & \textbf{95.17\%} & \textbf{0.87\%}\\  

\hline
\bottomrule
\end{tabularx}
\end{table*}}

\aptLtoX[graphic=no,type=html]{\begin{table*}
\caption{Dependency Coding Results}
\label{table:dep-2}
\begin{tabular}{ccccccc}
    \toprule
    {} &
    \multicolumn{3}{c|}{\textbf{Less Detail Prompts}} & 
    \multicolumn{3}{c}{\textbf{More Detail Prompts}}\\  
       \midrule
    \textbf{Total} & \multicolumn{3}{c|}{120} & 
    \multicolumn{3}{c}{112}\\
    \hline
    \textbf{Rating} & Correct \code{\textbf{(C)}} & Wrong \code{\textbf{(W)}} & Redundant \code{\textbf{(R)}}
  & Correct \code{\textbf{(C)}} & Wrong \code{\textbf{(W)}} & Redundant \code{\textbf{(R)}}\\
    \hline
    \textbf{Relationship} & \centering 89.17\% & 5.83\% & 5.00\% & 93.75\% & 1.79\% & 4.46\% \\
     \textbf{Mechanism} & 98.33\% & 1.67\% & n/a & 95.54\% & 4.46\% & n/a \\
    \bottomrule
\end{tabular}
\end{table*}}{\begin{table*}[h]
\caption{Dependency Coding Results}
\label{table:dep-2}
\small
\centering
\begin{tabularx}{\textwidth}{>{\centering\arraybackslash}X|>{\centering\arraybackslash}X>{\centering\arraybackslash}X>{\centering\arraybackslash}X|>{\centering\arraybackslash}X>{\centering\arraybackslash}X>{\centering\arraybackslash}X}
    \toprule
    {} &
    \multicolumn{3}{c|}{\textbf{Less Detail Prompts}} & 
    \multicolumn{3}{c}{\textbf{More Detail Prompts}}\\  
     \toprule
       \hline
    \textbf{Total} & \multicolumn{3}{c|}{120} & 
    \multicolumn{3}{c}{112}\\
    \hline
    \textbf{Rating} & Correct \code{\textbf{(C)}} & Wrong \code{\textbf{(W)}} & Redundant \code{\textbf{(R)}}
  & Correct \code{\textbf{(C)}} & Wrong \code{\textbf{(W)}} & Redundant \code{\textbf{(R)}}\\
    \hline
    \textbf{Relationship} & \centering 89.17\% & 5.83\% & 5.00\% & 93.75\% & 1.79\% & 4.46\% \\
     \textbf{Mechanism} & 98.33\% & 1.67\% & n/a & 95.54\% & 4.46\% & n/a \\
    \hline
    \bottomrule
\end{tabularx}
\end{table*}}

To evaluate the pipeline's ability to generate the task-driven data model based on user requests, we conducted a technical evaluation to assess the quality of the object-relational schema generated by LLMs, in our case, the GPT-4o model, which we used in our pipeline for corresponding modules. Specifically, we aim to assess:

\begin{itemize}
    \item \textbf{Schema Relevance}: whether the entities and attributes generated by the system are aligned with the user's task goals and contribute meaningfully to the task's completion.
    \item \textbf{Dependency Accuracy}: (a) the correctness of the relationships between entities, particularly whether the dependencies recognized by the system accurately model task-specific relationships; and (b) whether the corresponding mechanisms (i.e., update or validation) are correct. 
\end{itemize}

We did not evaluate the UI-specific aspects (e.g., UI components rendering and view composition mechanisms) in this technical evaluation, as these aspects depend heavily on user interaction and will be evaluated in the subsequent user study (Section~\ref{sec:user}).

\subsection{Setup}

\subsubsection{Dataset} 

We employed GPT-4o to generate a set of informational tasks that users might typically require interfaces to complete. The tasks spanned various domains to reflect the generalizability of the system across different informational needs. In total, our dataset comprises 25 task scenarios. To better understand the pipeline's ability to respond to prompts of different levels of detail, for each task, we generated two versions of task prompts---one less detailed and one more detailed, for example:

\begin{itemize}
    \item[-] Less Detailed Prompt: ``\textit{I want to plan a trip to Tokyo with my friends}.''
    \item[-] More Detailed Prompt: ``\textit{I want to plan a \underline{7-day} trip to Tokyo with my friends, \underline{for food and cultural experiences}}.''
\end{itemize}

This resulted in a dataset of 50 task requests, yielding a total of 197 entities, 1052 attributes, and 232 dependencies in all 50 corresponding data models\footnote{The data models used for evaluation are available on an interactive website: \href{https://jelly-modeleval.netlify.app/}{https://jelly-modeleval.netlify.app/}}.

\subsubsection{Coding Process} 
Two coders familiar with database schema and JavaScript programming (for analyzing dependencies) were involved in the evaluation. Each coder inspected the generated data models using the deployed data models and recorded their assessments on a coding sheet. For each task, the coders performed the following assessments:

\begin{itemize}
    \item \textbf{Schema Relevance:} Coders rated the relevance of each entity and attribute on a 4-point scale, ranging from (1) unreasonable (not relevant or redundant), (2) could be useful, (3) necessary and expected, to (4) useful and surprising.

    \item \textbf{Dependency Accuracy:} Coders assessed the dependency relationship among attributes and coded each as \textbf{\code{C}}orrect, \textbf{\code{W}}rong (e.g., missing or targeting incorrect attributes), or \textbf{\code{R}}edundant. Additionally, coders analyzed whether each dependency included correct validation or update relationship mechanism as described in Section~\ref{sec:dep} (i.e., correct JavaScript expression or natural language description) and coded each as either \textbf{\code{C}}orrect or \textbf{\code{W}}rong.
\end{itemize}

For any discrepancies between the codes of the dependency, a third coder examined the case to resolve the discrepancy. 

\subsection{Results}

As shown in Table~\ref{table:dep-1}, our coding results indicate that the majority of entities (94.12\% and 94.74\% for less and more detailed prompts) and attributes (93.91\% and 95.17\% for less and more detailed prompts) inferred by LLMs are necessary and expected.
These results demonstrate that the object-relational schema employed in our pipeline can effectively model users' tasks, providing relevant information tailored to their needs. In most cases, LLMs successfully generate meaningful entities and attributes. However, a common issue arises when LLMs interpret prompts too literally, leading to redundant attributes. For instance, in the task ``\textit{I want to buy a standing desk},'' the system might generate an unnecessary ``\textit{purchase decision}'' field. However, this only comprises less than 0.5\% of the cases.

Results of the dependency labeling in Table ~\ref{table:dep-2} show an average accuracy of 91.5\% for relationship and 96.9\% for mechanism. We identified two common errors: \textit{reversed relationships}, where the relationships reverse the source and target, and \textit{redundant dependencies}, where the dependencies describe relationships that are already declared in the schema with the referencing attributes. 
Although these dependency issues do not significantly impact the overall effectiveness of the pipeline and can often be identified through the GUI or corrected with improved prompting or validation, they highlight areas for improvement in how dependencies are inferred and generated. When using LLMs to establish directional relationships, validation is necessary to ensure accuracy, which we have integrated into our pipeline.


\section{User Study}
\label{sec:user}
We conducted a user study to gain a comprehensive understanding of the pipeline's capability in responding to real-world user requests, the effectiveness of the generated UIs, and the novel workflows and limitations that may emerge from the study. Specifically, we aim to answer the following research questions:

\begin{itemize}[leftmargin=2.5em]

\item [\textbf{RQ1}] Do the generated UIs effectively present information in a way that helps users accomplish their tasks?

\item [\textbf{RQ2}] How easily and effectively can users customize and modify the UIs to adapt to their evolving tasks and needs? 

\item [\textbf{RQ3}] What is the user experience of interacting with a generative and malleable information space, and how does it compare to existing interfaces such as chat-based systems and traditional GUI-based apps?

\item [\textbf{RQ4}] 
What challenges do users encounter when generating and adapting the UIs to suit their needs? Are these challenges primarily due to the limitations of the generation pipeline or the interaction mechanisms?

\end{itemize}

\subsection{Participants}
We recruited 8 participants (5 female and 3 male, aged 21--28) through the internal communication channels within a large public university. Participants, including students and research scientists, reported that they use diverse information systems in their daily life and work. All of them use generative AI tools (e.g., ChatGPT) on a daily basis. 

\subsection{Study Procedure}

The study lasted approximately 80 minutes per participant. Seven sessions were conducted in person and one remotely via Zoom. For all sessions, the experimenter provided verbal instructions, and the participants interacted with the system by following the experimenter's instructions.
All sessions were screen- and voice-recorded. Participants were provided with a consent form, which they reviewed and signed before the study. Participants received a 30 USD Amazon gift card for their participation. The study was divided into the following phases:

\textbf{Introduction (5 minutes).} Participants were first given a brief introduction to the research and an overview of the system, including the different parts of the interface and their functionality.

\textbf{System Walk-through/Tutorial (15 minutes).}  Participants were guided through an example task—hosting a dinner party, where they interacted with the system to request information, modify the UIs, and explore customization options. Participants performed the interactions themselves while the experimenter provided guidance, helping them familiarize themselves with the system's capabilities before moving on to freeform tasks.

\textbf{Freeform Tasks (2 tasks, each 20 minutes).} Participants were asked to use the system to complete two tasks of their choice. During each task, participants were encouraged to make various requests, evaluate the interface generated by the system, and reflect on whether the interface met their needs. They were instructed to create at least two follow-up queries per task to assess how well the system supported UI modifications.

\textbf{Questionnaire and Interview (20 minutes).}  After completing the tasks, participants filled out a 5-point Likert scale questionnaire evaluating different dimensions of the system and their overall experience. A semi-structured interview was also conducted to gather in-depth qualitative feedback.

\subsection{Results}

We summarize the results of the questionnaire, interview, and user behavior analysis for the free-form tasks.

\subsubsection{Questionnaire Results}
\label{sec:question-results}

Our questionnaire targeted assessing the utility and effectiveness of the generated information, the layout of the interface, and the interactions with \sys through prompting and direct manipulation (RQ1, RQ2). The results show that participants generally found the information presented on the interface relevant (6 strongly agree, 2 agree) and can help them achieve their tasks efficiently (2 strongly agree, 6 agree). Being able to express their personal needs and customize the interface accordingly is found to be easy (4 strongly agree, 4 agree) and useful (6 strongly agree, 2 agree). The panel layout and organization of information were found to be intuitive (6 strongly agree, 2 agree) and effective for information consumption (5 strongly agree, 2 agree, 1 neutral). Full questionnaire results can be found in Appendix~\ref{sec:questionnaire}.

\subsubsection{Open-Ended Tasks and Interaction Behaviors}
\begin{figure*}[t]
    \centering\includegraphics[width=\textwidth]{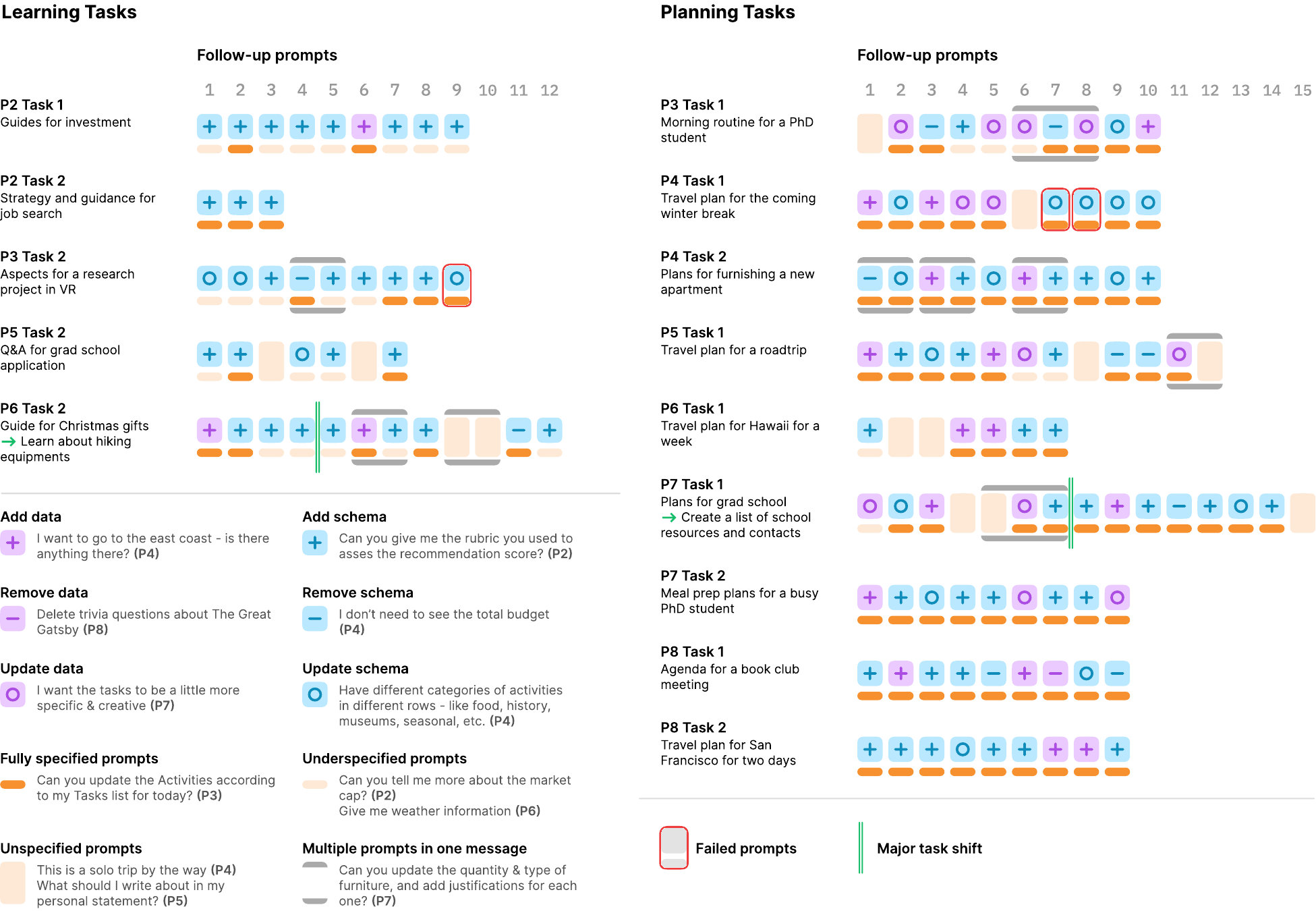}
    \caption[User study tasks and user prompts for continuous customization of the schema and data.]{User study tasks and user prompts for continuous customization of the schema and data, organized by types of free-form tasks users chose during the study, including sensemaking and brainstorming tasks (left) and planning and logistical tasks (right). Types of coding and examples for each type are included below the logs.
    }
    \label{fig:quantitative}
    \Description{The quantitative results of user study user prompts. Results for a total of 14 sessions are presented, where each prompt is visualized as a blue or purple rectangle, indicating a schema or data modification. The figure includes a two‐part chart shows Learning Tasks on the left and Planning Tasks on the right, each with multiple rows of tasks (labeled “P2 Task 1,” “P3 Task 2,” etc.) and a horizontal sequence of icons (pluses, minuses, circles, and shaded boxes). These icons represent different user follow‐up prompts such as adding or removing data, updating schemas, and so on. Some icons are outlined in red (indicating failed prompts), and vertical green lines mark major task shifts. A legend at the bottom shows examples of each prompt type (e.g., “Add data,” “Add schema”) along with the color or shape used in the chart. The overall figure illustrates how participants in a user study iteratively customized a schema and its data across a series of free‐form tasks.}
\end{figure*}

To better understand how users achieve their freeform tasks with the task-driven, model-based UIs (RQ3, RQ4), we logged and analyzed participants' follow-up chat messages with \sys for continuous customization of the interface (Fig.~\ref{fig:quantitative}).  We analyzed 14 out of 16 tasks (2 tasks missing due to the loss of P1's data), which included 120 follow-up messages. Occasionally, a single chat message contained multiple prompts with distinct requirements or pieces of information (see the example in Fig.~\ref{fig:quantitative}). We separated these prompts from the chat messages, which resulted in a total of 131 prompts. We include the full set of initial and follow-up prompts issued by participants to \sys in the supplementary materials.

We categorized the prompts participants initiated into \textit{Learning Task} and \textit{Planning Task}. The follow-up prompts are classified based on the intended updates of the data model---either the data or the schema---and the level of specificity. We visualize the logged results in Fig.~\ref{fig:quantitative}, with details reported below.

\textbf{Prompt Specificity.} We expect \sys to handle user prompts across varying levels of specificity. If a prompt specifies both \emph{target} and \emph{action} of the expected schema or data change (Section~\ref{sec:cont-prompt}), e.g., ``\textit{Add weather to the homepage}'' (P8), it is coded as \emph{fully specified}. If either or both aspects are missing, a prompt is considered \emph{underspecified}, e.g., ``\textit{Give me weather information}'' (P6). Sometimes, participants may broadly ask for strategies, e.g., ``\textit{What should I write about in my personal statement?}'' (P5); or simply provide contextual information, such as ``\textit{This is a solo trip by the way}'' (P4). These prompts do not provide any indication of how the UI should change. We code these prompts as \emph{unspecified} prompts.

Participants made a total of 91 \emph{fully specified} prompts (69\%), 27 \emph{underspecified} prompts (21\%), and 13 \emph{unspecified} prompts (10\%). Our results show that when participants prompt the system to adapt to their needs, they tend to be more specific on planning tasks (82\% \emph{fully specified}, 8\% \emph{underspecified}, 10\% \emph{unspecified}) than learning tasks (35\% \emph{fully specified}, 55\% \emph{underspecified}, 10\% \emph{unspecified}).

\textbf{Modification Patterns.} 
Participants issued a total of 118 \emph{fully specified} or \emph{underspecified} prompts to \sys to express the desired changes to either the schema or the data to adjust the UIs to their needs. Among these, 86 were schema changes (57 adds, 10 removes, 19 updates), and 32 were data changes (19 adds, 1 remove, 12 updates), indicating participants mostly sought to expand the scope of information or request additional data throughout the tasks.

Our results also show different modification patterns on the two types of tasks: during learning tasks, participants primarily engaged with schema modifications (92\% schema changes, 8\% data changes), whereas in planning tasks, the need for data modifications was significantly higher (65\% schema changes, 35\% data changes).

\textbf{Failure Cases.} While \sys was able to effectively interpret most user prompts and make corresponding schema and data changes, it occasionally failed during the study (a total of 3 times among 120 follow-up messages). We herein report these 3 failure cases:

\textit{Case 1:} \sys was unable to handle P3's request to customize the visual style of the UIs: ``\textit{Can you make the text size smaller?}'' As a result, no schema or data change was made.

\textit{Case 2:} P4 requested more acivities of different types to be displayed in separate rows, expecting \sys to create a set of lists, each containing a type of activity. Instead, \sys added a ``\textit{type}'' attribute to all existing activities, failing to meet the intended outcome.

\textit{Case 3:} After observing the error in Case 2, P4 repeated the same request. However, \sys maintained the same UI without implementing the desired changes.

These failure cases pointed to the limitations of the comprehensiveness of \sys's UI specifications, which we will discuss further in detail in Section~\ref{sec:eval-limitation}.

\subsection{Findings}
\label{sec:findings}

All participants expressed excitement about being able to generate an information space tailored to their needs. The iterative customization experience was perceived ``\textit{fun}'' (P2, P7) and ``\textit{efficient}'' (P2, P4), allowing them to have an interface that ``\textit{reflects their needs along the way}'' (P2). We discuss key findings and takeaways below.

\subsubsection{Effective Information Organization for Task Achievement [RQ1]}
\label{sec:rq2}

Our results, as discussed in Section~\ref{sec:question-results}, revealed participants found the structured information in \sys helped them effectively achieve their tasks.
Participants also noted that they appreciated the system's ability to extend beyond their direct prompts and generate ``\textit{reasonable surprises}'' that supported their goals (P4, P6). For example, when P4 requested new furniture for their living room planning task, \sys went beyond the prompt and grouped furniture based on aesthetic themes and suggested corresponding vendors. They were fond of the unexpected grouping of furniture into ``\textit{Bohemian}'' and ``\textit{Earth-toned}'' categories, noting how it mirrored their aesthetic preferences and saved their efforts in creating such groupings manually.
Moreover, \system commonly leveraged semantic linkages among those attributes with the proposed pipeline to streamline certain workflows. For example, when P7 requested dietary restrictions for all guests, \system not only applied those restrictions but also removed dishes that violated them, effectively anticipating the user's needs.

\textbf{\underline{Takeaway}}: The structured organization and semantic associations enabled by the object-relational schema effectively present and manage LLM's open-world knowledge to be easier to consume and control by the end-users. 



\subsubsection{Continuous Customizability and Flexible Adaptation to User Needs and Tasks [RQ2]}

One of the standout features of \system was its ability to accommodate continuous, iterative customization. P6 noted that many other tools offer ``\textit{one-shot}'' customization, where users make changes that are meant to be permanent. 
\sys's continuous adaptability allowed the users to adjust their workflows dynamically without being locked into a specific configuration. 
 For example, P8 noted that they often struggled to decide which applications to use and had to manually collect information from multiple sources into different note-taking tools. With \system, they could bypass the overhead of debating and selecting suitable applications as well as juggling multiple applications. They felt confident knowing they could always request \sys to provide the desired information as new needs arose. 

We observed participants naturally shifted focuses or changed task scopes when performing tasks with \system. For example, P6's task transformed from preparing Christmas gifts to learning and planning for a hiking trip, and P7 began with planning for settling into graduate school, which eventually scoped down to creating a list of school resources and managing their contacts. This fluidity was enabled by \system’s continuous customization, allowing participants to transition between evolving information needs without disrupting their current workflows. 


\sys's ability to interpret ambiguous prompts and leverage contextual information allowed participants to comfortably begin with vague inquiries and refine their goals as they interacted with the system. This was particularly beneficial when exploring unfamiliar domains. For example, during a trip to Hawaii, P6 stated, ``\textit{I want to stay on the beach}," without a clear idea of how this preference would impact their plans. They simply made the request out of curiosity to see how \sys would handle it. In response, \sys generated a list of beachfront hotels along with suggested beach activities, which inspired them to explore and plan different activities for the trip.

Participants also commented on the traceable history in the chat view, which allowed them to  easily revisit previous states, especially when AI-generated changes did not fully align with their needs.


\textbf{\underline{Takeaways:}} 
Users' intentions naturally shift and evolve during information tasks, varying in specificity. A system's ability to accommodate the varying levels of specificity and continuously adapt is essential for supporting fluid information activities. By leveraging LLMs' capacity to interpret flexible inputs and task-driven, model-based UI generation, and efficiently accomplish tasks.

\subsubsection{Task-Driven Data Model: Malleable but Persistent Structures [RQ3]}

Compared to chat-based systems and traditional apps, \system was seen as providing a novel, more fluid experience that combined the best of both worlds. The persistence and flexibility of the data model was considered beneficial in maintaining continuity with evolving tasks and able to offer desired interactions with the information they needed for the tasks (8 strongly agree).

Unlike existing chat-based interfaces with LLMs (e.g., ChatGPT), where responses primarily generate content, P6 noted that \system was ``\textit{generating the way to organize information}."
Participants also appreciated being able to make localized adjustments to a particular part of the interface---such as adding or editing a field---without needing to issue a full reset or restructuring of the entire layout, which is often the case in existing LLM-based generative systems (P2, P4). Moreover, the structure enabled users to modify and carry data seamlessly across ongoing tasks, reducing the need for copy-pasting and manual adjustments typically required in chat-based systems.
Transparency was another key advantage of the model-driven approach, giving participants confidence that their task structures remained intact rather than being unpredictably altered by AI.
As participants became more familiar with \system, they actively engaged with the schema view to explore and understand AI's modifications.
For example, when P6 requested daily weather information for their itinerary, they first checked the schema view to confirm that the itinerary schema had been extended before switching back to the UI view to verify the changes.

Compared to existing apps, participants appreciated the absence of ``\textit{opinionated}'' design choices that restrict customization (6 strongly agree, 1 agree, 1 neutral). P4 described situations where they could often achieve 80\% of their desired functionality in existing apps, but the inability to make small, yet necessary, modifications for specific use cases was frustrating---such as splitting road trip costs in a collaborative travel planning application. P6 further elaborated that existing apps ``\textit{force everyone to see all the information},'' 
whereas \system allowed users to view only what was relevant to them. This ability to "\textit{own my data}" and structure it according to personal preferences was seen as one of the major strengths (P6).


\textbf{\underline{Takeaways:}} Continuity and transparency are essential for a generated interface to effectively meet users' task needs. Anchoring the generation process to the flexible, task-driven data model ensures these qualities, enabling users to adapt structures to their preferences and maintain confidence in the UI modifications.

\subsubsection{Expecting More Efficient Ways to Interact with the Model and UIs}
\label{sec:eval-limitation}

One challenge of the current implementation of \system was the need for continuous prompting to make incremental changes for most of the cases. 
 While all participants acknowledged that it was easy to articulate intended changes to AI (4 strongly agree, 4 agree),
P1 and P8 noted that it could be tedious to describe every requirement in detail when the system failed to initially generate a sufficient task structure. A potential improvement is to generate the model along with possible expansions to it (e.g., additional entities and attributes to consider), which would enable users to quickly expand the model with lightweight interaction.

Three participants (P3, P6, and P7) found the schema view useful for inspecting the underlying structure and understanding the changes, while others mainly relied on the generated UI for their tasks. 
P4 suggested that enhancing interactions for manipulating the schema directly could be particularly beneficial for developers.
For example, P4 expressed a desire to ``\textit{cherry-pick}'' elements from one schema as the starting point for another session. This also points to a potential future direction for making schema composable and reusable to enable users to create new spaces from existing ones, facilitating cross-domain tasks.

Additionally, the generated UIs in \system incorporated a limited set of design patterns within its specifications, restricting the expressiveness of information presentation and interactions. For example, P2 expressed the desire to have a line chart visualization of stock information, which was beyond the scope of our system's current UI specifications. This limitation highlights the need for a more comprehensive UI specification to enhance information representation and interaction (e.g., diverse layouts, advanced interaction logic), which we discuss further in the following section.

\textbf{\underline{Takeaways: }} 
Efficiency of information acquisition and expressiveness of UIs are desired by the users. While continuous prompting offers flexibility, more proactive model expansion and lightweight refinement mechanisms are needed for achieving composable and reusable UIs to further adapt to evolving task needs. Enhancing UI design patterns and expanding UI specifications would further improve information expressiveness and usability.

\section{Discussion and Future Directions on Generative and Malleable UI}
\label{sec:discussion}
In this section, we reflect on our findings, their implications, and future directions for advancing generative and malleable UIs. We revisit the five key aspects outlined in Section~\ref{sec: aspects}, structuring our discussion around them.


\vspace{-3pt}

\subsection{Broadening Supported Tasks with Advanced Dependency Modeling}
Our technical pipeline currently models the dependencies required in an information task by describing the relationships among pairs of source and target elements. While this relatively simple mechanism yields few errors in our technical evaluation and is sufficient in supporting study participants in completing their intended tasks, it also has limitations in describing tasks that require complex interaction. A future direction is to explore more advanced graph-based dependency modeling, where nodes of the graph represent entities and attributes, and the edges represent the dependency relationships expressed using a more expressive specification language. This dependency graph will not only allow for the modeling of interaction logic beyond pairs of elements but also will enable end-users to intuitively inspect the dependencies by leveraging representation and interaction techniques introduced in graph-based visual programming.

\vspace{-3pt}

\subsection{Supporting Information Transformation Patterns with Higher-level Schema Operations}
While the current schema operations (e.g., add, delete, and update) could theoretically handle all possible schema transformations, having to translate high-level transformations to these low-level operations can lead to complex and error-prone data and UI modifications---especially when LLMs are involved in the process. High-level transformations, such as \textit{eversion}, are commonly needed but difficult to express with atomic operations. For example, a user might start by viewing a list of literature modeled as an \texttt{Publication} entity containing attributes like title, authors, and year. Later, they might wish to transform this view to focus on all \texttt{Authors} and their respective publications. Using current schema operations, the system would need to generate a new entity, likely resulting in substantial changes to both the UI and underlying data. A dedicated \textit{eversion} schema operation from the \texttt{Publication} entity to the \texttt{Author} entity could ensure a smooth data and UI transformation. A future direction is to analyze users' prompts with the system to identify the desired high-level transformations and expand the schema operations to support them more directly.

\vspace{-3pt}

\subsection{Enabling Advanced and Malleable View Management}
The current \sys implementation employs a column-based layout to organize the panels on the screen. We intentionally opted for this layout as it avoids the tedious and manual positioning of the panels while ensuring sufficient usability of the generated UIs. Its simplicity allows us to focus our research on investigating the underlying technical pipeline. View and layout management in itself is a long-standing research theme in HCI~\cite{pastpresentfutureUI, bach2022dashboard, zeidler2013aucklandlayout, interactivespecificationLayout}. Among the various methods, a straightforward extension of the \sys system is to integrate the underlying data model with the dashboard design pattern, which summarizes the key dimensions that guide the placement and presentation of information panels~\cite{bach2022dashboard}. 
Concretely, entities and lists of entities can be displayed in separate panels, arranged based on inferred importance—such as the number of attributes or connections to other entities. Key panels could occupy central positions with detailed information, whereas less critical ones might be placed along the periphery with more condensed views. Certain entities may benefit from being displayed with multiple synchronized representations (e.g., a table and a map).
By supporting the diverse representations of information within each panel, we can further enhance \sys's expressivity in the graphical representation of information.

\vspace{-3pt}

\subsection{Integrating with External and User Data and Enhanced Data Transformation and Management}
\label{sec:data}

An immediate next step is to extend our technical pipeline with external data sources and user-permitted data beyond just relying on LLM-generated data. Recent approaches of integrating LLMs with external data---such as Retrieval-Augmented Generation \cite{lewis2020retrieval},  the Model-Context Protocol \cite{anthropic_model_context_protocol}, and LLM-generated API calls \cite{trivedi2024appworld}---offer promising ways of integrating reliable data sources into our technical pipeline. In addition to connecting with the various data sources, it is also worth systematically exploring the underlying data infrastructure with regard to how data is managed and transformed. The benefit of a persistent schema employed by traditional applications is that it may be optimized based on data storage and retrieval efficiency. The dynamic data schema that drives the interface may not directly mapped to the underlying database schema. Our future work will investigate a data transformation layer to facilitate seamless acquisition and data adaptation in response to evolving schema changes.

\vspace{-3pt}

\subsection{Enabling Personalized UIs with Context Preservation}
Personalized interfaces have been a long-standing endeavor in activity-centered computing~\cite{adar1999haystack,karger1997haystack}, which require not only customization but also context preservation. Unlike most existing interfaces that remain the same regardless of who uses them and for how long, we will explore how the underlying data model and the generated UIs can become personalized and intelligently tailored to each individual context and preference over time. 
A key challenge is balancing adaptability and predictability. An interface that adapts too aggressively may make incorrect assumptions about user intent, leading to frustration; conversely, an interface that requires excessive manual configurations, shifts too much burden on the user. 
Our insight is that achieving context-aware UIs requires an intermediate representation that effectively preserves and reuses user context and hence guides adaptation.

The data model presents a promising solution. \sys can record users' preferred entities, attributes, and interface configurations for different tasks. When a user encounters new tasks, it can intelligently reuse subcomponents from previous relevant tasks. For example, if a user is organizing an academic workshop for the first time, the system could adapt UI elements from workspaces of past activities, such as scheduling talks (as in conference planning) or coordinating a dinner event (as in family gathering). 
Besides model reuse, future work will also explore personalized model evolution. Different users may prioritize different aspects of their workflows---a researcher may focus on entity relationships when conducting a literature review, while a project manager may emphasize task dependencies and deadlines. 
Through the interactions we explore with the data model, users should be able to inspect and customize the context being preserved and adapted, achieving an information space that is not only generative, malleable, but also personal. 

\section{Conclusion}
\label{sec:conclusion}
As users can dynamically express arbitrary prompts to AI describing their intended information activity, a demand arises for generative and malleable user interfaces. This poses an enormous and exciting challenge for the HCI community---how can we design interfaces that can support information activities of any domain, scope, and complexity? This work provides an exploration towards this goal.
Taking the perspective that a GUI-based interactive system is the graphical representation of the data model that describes the targeted user tasks, we recognize that generative and malleable user interfaces fundamentally demand generative and malleable data models to support users' dynamic tasks. Therefore, we propose leveraging LLMs to generate task-driven data models based on the tasks indicated in users' prompts, which then guide the generation of the user interface. Results from our technical evaluation show that LLMs can generate relatively high quality data models. We implemented the proposed generation pipeline into a prototype system \sys. The user evaluation of the system shows that the generative and malleable user interfaces enable users to develop a personalized and dynamic information space by flexibly curating diverse information and customizing its representation. Our research demonstrates a promising approach and points out a wide range of exciting future research directions to realize the long-envisioned generative, malleable user interfaces.

\bibliographystyle{ACM-Reference-Format}

\appendix

\clearpage
\onecolumn

\newcommand{\keycode}[1]{\code{#1}}
\newcommand{\valuecode}[1]{\code{#1}}

\section{UI Specification}
\label{sec:specs}

\aptLtoX[graphic=no,type=html]{\begin{table}
\caption{UI specification for one attribute of an object}
\begin{tabularx}{\textwidth}{@{}ccX@{}}
\toprule
\textbf{Key} & \textbf{Value} & \multicolumn{1}{c}{\textbf{Explanation}} \\ \midrule
\multirow{4}{*}[-0.6em]{\textbf{\code{type}}} & \code{string} & \multirow{4}{*}[-0.6em]{\begin{tabularx}{\linewidth}{X}The data type of the value, which can be a string, number, array, or the name of an entity, for which we use \valuecode{\_\_\textless{}ENTITY\textgreater{}\_\_} to annotate, e.g., \valuecode{\_\_PERSON\_\_}\end{tabularx}} \\ \cline{2-2}
 & \code{number} & \\ \cline{2-2}
 & \code{array} & \\ \cline{2-2}
 & \code{\_\_\textless{}ENTITY\textgreater{}\_\_} & \\ \hline
\multirow{3}{*}[-2em]{\textbf{\code{function}}} & \multirow{1}{*}[-0.6em]{\code{priviateIdentifier}} & The attribute functions as a unique identifier of an object used internally by the \sys system. Private identifiers may not be semantically meaningful to the user and should not be displayed on the interface \\ \cline{2-3}
 & \multirow{1}{*}[-0.6em]{\code{publicIdentifier}} & The attribute functions as a representative identity of an object, such as name and title. Public identifiers should be displayed with the highest saliency when rendering the UI for the object \\ \cline{2-3}
 & \code{display} & All other attributes of the object \\ \hline
\textbf{\code{editable}} & \code{true} or \code{false} & Whether the user is allowed to modify the value of the attribute in the rendered UI \\ \hline
\multicolumn{3}{c}{\textbf{When the {\keycode{type}} of the attribute is \underline{not} {\valuecode{array}} ...}} \\ \hline
\multirow{8}{*}[-1.75em]{\textbf{\code{render}}} & \code{shortText} & Short pieces of text, e.g., name of a hotel, title of a book \\ \cline{2-3}
 & \code{paragraph} & Long blocks of text, e.g., description of a city, review of a product \\ \cline{2-3}
 & \code{number} & Numeric values, including integers, floats, percentages, etc. \\ \cline{2-3}
 & \code{url} & Links to websites, e.g., \url{https://hci.ucsd.edu} \\ \cline{2-3}
 & \code{time} & Temporal values, such as dates, specific points in time, durations, etc. \\ \cline{2-3}
 & \code{location} & Geographic coordinates or names of places \\ \cline{2-3}
 & \code{category} & One of the categories defined by the \code{categories} field as a list of strings in the attribute specification \\ \cline{2-3}
 & \code{hidden} & The attribute will not be rendered \\ \hline
\multicolumn{3}{c}{\textbf{When the {\keycode{type}} of the attribute is {\valuecode{array}} ...}} \\ \hline
\multirow{2}{*}[-1em]{\textbf{\code{render}}} & \multirow{1}{*}[-0.6em]{\code{summary}} & The array is minimized as a single button showing one key aspect of the items in the array (see below for how it's derived). The full array shows upon clicked \\ \cline{2-3}
 & \code{expanded} &The array is fully shown, and the user can directly see, scroll through, and interact with each item \\ \hline
\multirow{2}{*}[-1em]{\textbf{\code{item}}} & \code{type} & The type of the items in the array, same as \keycode{type} for the attributes \\ \cline{2-3}
 & \multirow{1}{*}[-0.6em]{\code{thumbnail}} & An array of attribute names. If the item type is an entity, we need a set of representative and relevant attributes when displayed in a minimized form in the array (default to \valuecode{publicIdentifier}) \\ \hline
\multirow{1}{*}[-1.3em]{\textbf{(\code{summary})}} & \multirow{1}{*}[-1.3em]{\code{\{ name, derived \}}} & Only for the \valuecode{summary} render type. The target attribute of the array items (\keycode{name}) and the method for deriving the summarizing value from them (\keycode{derived}). \keycode{derived} is an object of two keys, \keycode{field} and \keycode{operation}. If \keycode{field} is a number, \keycode{operation} could be \valuecode{SUM}, \valuecode{AVG}, \valuecode{MIN}, or \valuecode{MAX}; or \valuecode{FILTER} or \valuecode{COUNT} if \keycode{field} is an array \\ \bottomrule
\end{tabularx}
\end{table}}{\begin{table}[h]
\renewcommand{\arraystretch}{1.5}
\caption{UI specification for one attribute of an object}
\small
\begin{tabularx}{\textwidth}{@{}ccX@{}}
\toprule
\hline
\addlinespace[0.5em]
\textbf{Key} & \textbf{Value} & \multicolumn{1}{c}{\textbf{Explanation}} \\ \addlinespace[0.5em]\midrule
\multirow{4}{*}[-0.6em]{\textbf{\code{type}}} & \code{string} & \multirow{4}{*}[-0.6em]{\begin{tabularx}{\linewidth}{X}The data type of the value, which can be a string, number, array, or the name of an entity, for which we use \valuecode{\_\_\textless{}ENTITY\textgreater{}\_\_} to annotate, e.g., \valuecode{\_\_PERSON\_\_}\end{tabularx}} \\ \cmidrule(lr){2-2}
 & \code{number} & \\ \cmidrule(lr){2-2}
 & \code{array} & \\ \cmidrule(lr){2-2}
 & \code{\_\_\textless{}ENTITY\textgreater{}\_\_} & \\ \midrule
\multirow{3}{*}[-2em]{\textbf{\code{function}}} & \multirow{1}{*}[-0.6em]{\code{priviateIdentifier}} & The attribute functions as a unique identifier of an object used internally by the \sys system. Private identifiers may not be semantically meaningful to the user and should not be displayed in the interface \\ \cmidrule(l){2-3} 
 & \multirow{1}{*}[-0.6em]{\code{publicIdentifier}} & The attribute functions as a representative identity of an object, such as name and title. Public identifiers should be displayed with the highest saliency when rendering the UI for the object \\ \cmidrule(l){2-3} 
 & \code{display} & All other attributes of the object \\ \midrule
\textbf{\code{editable}} & \code{true} or \code{false} & Whether the user is allowed to modify the value of the attribute in the rendered UI \\ \midrule\hline
\addlinespace[0.25em]
\multicolumn{3}{c}{\textbf{When the {\keycode{type}} of the attribute is \underline{not} {\valuecode{array}} ...}} \\ \addlinespace[0.25em]\midrule
\multirow{8}{*}[-1.75em]{\textbf{\code{render}}} & \code{shortText} & Short pieces of text, e.g., name of a hotel, title of a book \\ \cmidrule(l){2-3} 
 & \code{paragraph} & Long blocks of text, e.g., description of a city, review of a product \\ \cmidrule(l){2-3} 
 & \code{number} & Numeric values, including integers, floats, percentages, etc. \\ \cmidrule(l){2-3} 
 & \code{url} & Links to websites, e.g., \url{https://hci.ucsd.edu} \\ \cmidrule(l){2-3} 
 & \code{time} & Temporal values, such as dates, specific points in time, durations, etc. \\ \cmidrule(l){2-3} 
 & \code{location} & Geographic coordinates or names of places \\ \cmidrule(l){2-3} 
 & \code{category} & One of the categories defined by the \code{categories} field as a list of strings in the attribute specification \\ \cmidrule(l){2-3} 
 & \code{hidden} & The attribute will not be rendered \\ \midrule\hline
 \addlinespace[0.25em]
\multicolumn{3}{c}{\textbf{When the {\keycode{type}} of the attribute is {\valuecode{array}} ...}} \\ \addlinespace[0.25em]\midrule
\multirow{2}{*}[-1em]{\textbf{\code{render}}} & \multirow{1}{*}[-0.6em]{\code{summary}} & The array is minimized as a single button showing one key aspect of the items in the array (see below for how it's derived). The full array shows upon clicked \\ \cmidrule(l){2-3} 
 & \code{expanded} &The array is fully shown, and the user can directly see, scroll through, and interact with each item \\ \midrule
\multirow{2}{*}[-1em]{\textbf{\code{item}}} & \code{type} & The type of the items in the array, same as \keycode{type} for the attributes \\ \cmidrule(l){2-3} 
 & \multirow{1}{*}[-0.6em]{\code{thumbnail}} & An array of attribute names. If the item type is an entity, we need a set of representative and relevant attributes when displayed in a minimized form in the array (default to \valuecode{publicIdentifier}) \\ \midrule
\multirow{1}{*}[-1.3em]{\textbf{(\code{summary})}} & \multirow{1}{*}[-1.3em]{\code{\{ name, derived \}}} & Only for the \valuecode{summary} render type. The target attribute of the array items (\keycode{name}) and the method for deriving the summarizing value from them (\keycode{derived}). \keycode{derived} is an object of two keys, \keycode{field} and \keycode{operation}. If \keycode{field} is a number, \keycode{operation} could be \valuecode{SUM}, \valuecode{AVG}, \valuecode{MIN}, or \valuecode{MAX}; or \valuecode{FILTER} or \valuecode{COUNT} if \keycode{field} is an array \\ \addlinespace[0.5em]\hline\bottomrule
\end{tabularx}
\end{table}}

\clearpage

\subsection{Example Specification}

\begin{lstlisting}[language=json,firstnumber=1]
{
  DINNER_PLAN: {
    id: { type: "string", editable: false, render: "hidden", function: "privateIdentifier" },
    date: { type: "string", editable: true, render: "date", function: "display" },
    host: { type: "__USER__", editable: true, render: "shortText", function: "display" },
    location: { type: "string", editable: true, render: "location", function: "display" },
    guest_list: {
      type: "array",
      editable: true,
      render: "expanded",
      item: { type: "__USER__", thumbnail: ["name", "phone"] }
    },
    menu: {
      type: "array",
      editable: true, 
      render: "summary",
      summary: {
        name: "total_calories",
        derived: { operation: "SUM", field: "calories" }
      },
      item: { type: "__DISH__", thumbnail: ["name", "calories"] }
    }
  },
  USER: {
    id: { type: "string", editable: false, render: "hidden", function: "privateIdentifier" },
    name: { type: "string", editable: false, render: "shortText", function: "publicIdentifier" },
    email: { type: "string", editable: true, render: "url", function: "display" },
    phone: { type: "string", editable: true, render: "number", function: "display" }
  },
  DISH: {
    id: { type: "string", editable: false, render: "hidden", function: "privateIdentifier" },
    name: { type: "string", editable: true, render: "shortText", function: "publicIdentifier" },
    ingredients: {
      type: "array",
      editable: false,
      render: "expanded",
      item: { type: "string", editable: false, render: "shortText", function: "display" }
    },
    calories: { type: "number", editable: false, render: "number", function: "display" },
    cuisine_type: {
      type: "string",
      editable: false,
      render: "category", 
      function: "display",
      categories: ["American", "Italian", "Chinese", "Japanese", "French"]
    }
  }
}
\end{lstlisting}

\clearpage

\section{User Study Questionnaire and Responses}
\label{sec:questionnaire}

We plot the summary of all 5-point Likert scale questionnaire questions and participant responses here.

\begin{figure*}[ht]
    \centering
    \includegraphics[width=0.67\textwidth]{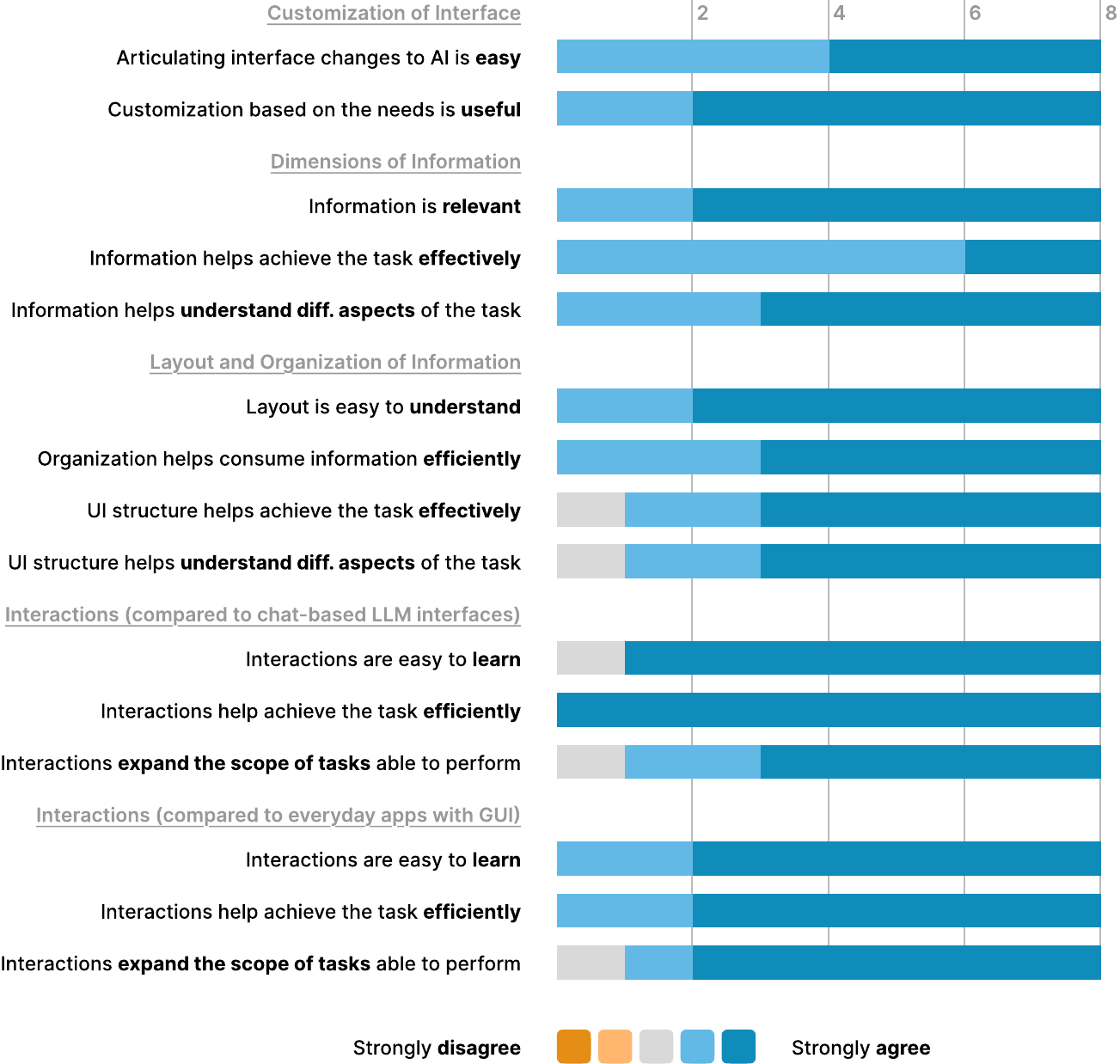}
    \caption{5-point Likert scale questionnaire questions and user responses.}
    \label{fig:questions}
    \Description{A horizontal bar chart displays participant ratings (from strongly disagree to strongly agree, in color gradients) across multiple statements grouped under headings: Customization of Interface, Dimensions of Information, Layout and Organization of Information, and Interactions (compared to chat-based LLMs and everyday GUI apps). Each row on the left is a statement (e.g., “Articulating interface changes to AI is easy,” “Information is relevant,” “Interactions are easy to learn”), with a horizontal stacked bar to the right showing how many participants selected each Likert response. Most bars show predominantly darker blues on the right, indicating that participants generally agreed or strongly agreed with the statements.}
\end{figure*}

\end{document}